\documentclass[journal=jctcce,manuscript=article,layout=twocolumn]{achemso}
\setkeys{acs}{maxauthors=10,etalmode=truncate,doi=true}
\renewcommand{\suppinfoname}{Supporting Information}

\usepackage{graphicx}
\usepackage{booktabs}
\usepackage{enumitem}

\usepackage{float}
\usepackage[section]{placeins}
\usepackage{stfloats}

\usepackage{amsmath}
\usepackage[version=4]{mhchem}
\usepackage{siunitx}

\usepackage{xcolor}
\usepackage{soul}
\sethlcolor{yellow}

\definecolor{mvdblue}{HTML}{356FA8}
\usepackage{hyperref}
\hypersetup{colorlinks=true,allcolors=mvdblue,pdfborder={0 0 0}}
\newcommand{\namedref}[2]{\hyperref[#2]{#1~\ref*{#2}}}
\newcommand{\secref}[1]{\namedref{Section}{#1}}
\newcommand{\figref}[1]{\namedref{Figure}{#1}}
\newcommand{\figpanelref}[2]{\hyperref[#1]{Figure~\ref*{#1}#2}}
\newcommand{\tabref}[1]{\namedref{Table}{#1}}
\newcommand{\eqnref}[1]{\namedref{eq}{#1}}

\newcommand{\figrefs}[2]{\namedref{Figures}{#1} and \hyperref[#2]{\ref*{#2}}}

\makeatletter
\def\doi@aux#1{\endgroup\href{https://doi.org/#1}{\nolinkurl{#1}}}
\makeatother

\DeclareSIUnit\angstrom{\text{Å}}
\DeclareSIUnit\kcal{kcal}
\DeclareSIUnit\bar{bar}

\title{Choosing Spatial Resolution in the Residence-Time Approach to Local Diffusivity}
\author{Michael von Domaros}
\affiliation[UMR]{Fachbereich Chemie, Philipps-Universität Marburg, 35032 Marburg, Germany}
\email{mvondomaros@uni-marburg.de}
\author{Rinto Thomas}
\affiliation[UMR]{Fachbereich Chemie, Philipps-Universität Marburg, 35032 Marburg, Germany}
\altaffiliation{R.T. and P.R.P. contributed equally to this work.}
\author{Praveen~Ranganath~Prabhakar}
\affiliation[UCI]{Department of Chemistry, University of California, Irvine, Irvine, California 92697, United States}
\alsoaffiliation{Present address: Department of Chemistry, University of Michigan, Ann Arbor, Michigan 48109, United States}
\altaffiliation{R.T. and P.R.P. contributed equally to this work.}

\usepackage{needspace,threeparttable}
\begin{document}
\maketitle

\begin{abstract}
	Estimating position-dependent diffusivity with the residence-time approach
	(RTA) requires choosing an interval width that resolves spatial variation
	while probing time scales compatible with a local diffusion model.
	We develop a selection criterion that combines bulk diffusivity convergence
	with local model agreement measured by the Kullback--Leibler divergence
	between observed and predicted first-exit-time distributions, and assess
	global model agreement through independent propagator analysis against
	unbiased molecular dynamics.
	Applications to water in a fluid phospholipid bilayer and three permeants
	in a model skin lipid barrier show that local agreement alone does not
	ensure global fidelity: in the skin barrier, narrow intervals minimize
	exit-time mismatch but predict excessive spreading, whereas requiring bulk
	convergence selects wider intervals with better longer-time predictions.
	In propagator tests, RTA profiles at the selected interval widths outperform
	velocity-autocorrelation-based profiles and generally improve agreement
	relative to position-autocorrelation-based profiles in the skin barrier,
	while phospholipid-bilayer rankings depend on lag time and discrepancy
	metric; both autocorrelation methods use harmonically restrained simulations.
\end{abstract}

\section{Introduction}

In a homogeneous liquid, the self-diffusion coefficient $D$ characterizes
the long-time growth of a molecule's mean-square displacement.
In an inhomogeneous medium, molecular mobility varies with the local
environment: a molecule may move readily in one region but be strongly
hindered in another.\cite{LiuHarderBerne2004,MittalTruskett2008} A position-dependent
diffusion coefficient $D(x)$, with $x$ denoting a spatial coordinate,
describes this variation, with larger values corresponding to faster
local diffusive spreading. On time and length scales for which a
Smoluchowski description is appropriate,\cite{Hummer2005,LauLubensky2007}
the probability density $p(x,t)$ of the molecule's position evolves in one
dimension according to
\begin{equation}
	\begin{aligned}
		\partial_t p(x,t)
		 & =\partial_x\!\Bigl[D(x)\Bigl(\partial_x p(x,t)\\
			                &\qquad+\beta p(x,t)\,\partial_x F(x)\Bigr)\Bigr],
	\end{aligned}
	\label{eq:smoluchowski}
\end{equation}
where $F(x)$ is the potential of mean force (PMF) and
$\beta=1/(k_{\mathrm B}T)$, with Boltzmann's constant $k_{\mathrm B}$ and
temperature $T$. The PMF describes the molecule's thermodynamic
preferences, while the diffusion coefficient describes its local
mobility; together, they determine its dynamics within this model.

Lipid membranes provide a well-studied example of an inhomogeneous medium:
a permeant encounters distinct molecular environments as it moves from
the surrounding aqueous phase through the membrane interface and into the lipid
interior.\cite{MarrinkBerendsen1994,GhyselsVenable2017} The inhomogeneous
solubility--diffusion (ISD) model connects the profiles $F(x)$ and $D(x)$
along the membrane normal to the steady-state permeability $P$. For dilute
permeants, $P$ is the proportionality constant between the steady flux and
the bulk concentration difference across the membrane. Within this model,
its inverse is the integrated resistance to
permeation:\cite{Diamond1974Partition,Venable2019Permeability}
\begin{equation}
	P^{-1}=\int_{x_a}^{x_b}
	\frac{e^{\beta[F(x)-F_{\mathrm{bulk}}]}}{D(x)}\,\mathrm dx,
	\label{eq:isd}
\end{equation}
where $F_{\mathrm{bulk}}$ is the PMF in the bulk reference phase, and $x_a$
and $x_b$ delimit the transport region. The integrand gives the local
resistance per unit length, identifying the regions that contribute most
strongly to the total resistance.

When permeability is the sole quantity of interest, counting spontaneous
crossings can provide a direct estimate if sufficiently many events are
observed.\cite{KramerGhysels2020} Rare-event path methods provide an
alternative route to permeation rates when spontaneous crossings are
difficult to sample.\cite{GhyselsRoet2021,VervustZhang2023} Neither route
requires a local diffusion model. Such a model, however, provides insight
beyond the overall permeability: within the ISD description, $F(x)$ and
$D(x)$ distinguish resistance arising from unfavorable occupancy from
resistance arising from low mobility. Diffusivity profiles also support
spatially resolved predictions of molecular motion that can be tested
against trajectory data.\cite{DeVosVenable2018}

Estimating $D(x)$ from molecular dynamics (MD) requires choosing the time
scale on which a diffusion model represents molecular motion, often
through an explicit lag time. Available methods infer diffusivity from
position, velocity, and force autocorrelation functions measured in
restrained or constrained simulations,\cite{WoolfRoux1994,Hummer2005,MarrinkBerendsen1994,GaalswykAwoonorWilliams2016}
from local displacement statistics,\cite{Nagai2020LocalMSD,Kikkawa2026LocalDiffusion}
or by fitting diffusion equations or Markov state models to transition
statistics.\cite{ComerChipot2013,SicardKoskin2021} Recent work also infers
diffusivity by reconstructing time-dependent friction kernels from
correlation functions measured in restrained simulations.\cite{HarrisRoux2026}
These methods require choices, such as spatial discretization, restraint force
constants, or fitting parameters, whose effects on the inferred
diffusivity must be assessed.

Restraints can alter the dynamics themselves:
strong harmonic potentials, for example, can substantially increase the
solvent friction on a molecular solute.\cite{Daldrop2017ExternalPotential}
Even without harmonic restraints, molecular motion can retain memory and exhibit
subdiffusion over extended time ranges.\cite{ChipotComer2016,HarrisChipotRoux2024}
A Markovian diffusion model assumes that the current position alone
determines the statistics of subsequent motion and may therefore be
inadequate on these time scales. Estimating a diffusivity profile must
thus be accompanied by testing whether the resulting model reproduces
the observed dynamics on the time and length scales of interest.

Exit-time statistics offer a physically intuitive route to diffusivity:
the time a molecule takes to leave a spatial region reflects its mobility
within that region.\cite{Redner2001FirstPassage,MercierFrancoCastier2016,HollringBaer2023}
Our recently introduced residence-time approach (RTA) estimates local
diffusivity from the mean first-exit time averaged over starting positions
within an interval, assuming a flat effective
PMF.\cite{Thomas2026RTA} This mean, termed the residence time, is the only
statistic required by the estimator.

The interval width $L$ is its sole
adjustable analysis parameter and sets both the spatial resolution and,
through the characteristic exit times, the time scale probed. Narrow
intervals resolve spatial variation more finely, but their short exit
times may probe motion that retains memory. Wider intervals probe longer times, over
which memory effects may decay, but average over more spatial variation.
Choosing $L$ therefore requires balancing spatial resolution against the
time scales on which local diffusion adequately describes molecular motion.

Here we develop a reproducible criterion for selecting the interval width
$L$ using bulk diffusivity convergence and agreement between observed
exit-time statistics and a constant-diffusivity model. We first describe
the RTA estimator and the measures of local and global model agreement
in \secref{sec:theory}, followed by the simulation and analysis procedures
in \secref{sec:computational-details}. In \secref{sec:results}, we examine
width dependence, formulate the selection criterion using a fluid lipid
bilayer, and apply it to a model of the skin's lipid barrier. We then
test the profiles at the selected interval widths against independent,
unbiased trajectories, assessing their predictions beyond the exit-time
statistics used for width selection.

\section{Theory}

\label{sec:theory}

We first relate the mean first-exit time to the RTA diffusivity estimate.
We then define two complementary tests of the diffusion model: local
agreement with exit-time distributions and global agreement with
propagators from unbiased MD.

\subsection{Residence-time approach}

RTA assumes a flat effective PMF within each residence
interval. Any bias potential $V_{\mathrm b}(x)$ that cancels the spatial
variation of $F(x)$ gives
$F_{\mathrm{eff}}(x)=F(x)+V_{\mathrm b}(x)=\mathrm{const}$.
In practice, we use adaptive biasing force (ABF) sampling to approach this
limit.\cite{Darve2001ABF,ComerGumbart2015}
For biased dynamics in a flat effective PMF,
\eqnref{eq:smoluchowski} reduces to
\begin{equation}
	\partial_t p(x,t)
	=\partial_x\!\left[D(x)\,\partial_x p(x,t)\right].
\end{equation}

For an interval of width $L$ centered at $x$, RTA further assumes constant
diffusivity $D$ and Markovian motion on the time scales probed by first exits.
Within each interval, the model neglects spatial variation in diffusivity.

The first-exit time $T$ is the time a particle
starting inside the interval takes to reach either boundary. We define
the residence time $\tau$ as the mean of $T$ over trajectories and starting
positions. Under a flat effective PMF, the stationary distribution of
starting positions is uniform. For this distribution, the local diffusion
model gives $\tau=\langle T\rangle=L^2/(12D)$,\cite{Redner2001FirstPassage}
yielding the estimator
\begin{equation}
	\widehat D(x;L)=\frac{L^2}{12\widehat\tau(x;L)},
	\label{eq:rta}
\end{equation}
where $\widehat\tau(x;L)$ is the sample mean of first-exit times measured
in the biased MD trajectory. Repeating the analysis for intervals centered
at different positions gives the diffusivity profile.

\subsection{Local model agreement}

The RTA estimate $\widehat D$ reproduces the measured residence time by
construction. To test the model beyond this mean, we compare the observed
first-exit-time distribution in each residence interval with the
analytical prediction at $D=\widehat D$, assuming constant diffusivity
and a flat effective PMF. We express this comparison as an excess
negative log-likelihood per observation so that model agreement can be
compared across interval widths.

The survival probability $S_L(t;D)=\Pr(T>t)$ is the probability that a
particle has not reached either boundary by time $t$. For constant
diffusivity $D$ and uniformly distributed starting positions in an
interval of width $L$, the corresponding flat-PMF model gives\cite{Redner2001FirstPassage}
\begin{equation}
	\begin{aligned}
		S_L(t;D)
		 & =\frac{8}{\pi^2}\sum_{m=0}^{\infty}
		\frac{e^{-\lambda_m t}}{(2m+1)^2},     \\
		\lambda_m
		 & =\frac{(2m+1)^2\pi^2 D}{L^2}.
	\end{aligned}
	\label{eq:rta-survival}
\end{equation}
where $\lambda_m$ are the decay rates of the diffusion eigenmodes.

Let $t_k=k\Delta t$, where $\Delta t$ is the time between saved MD frames.
The model probability of a first exit during $(t_{k-1},t_k]$ is
$q_k=S_L(t_{k-1};\widehat D)-S_L(t_k;\widehat D)$.
If $n_k$ is the number of observed first exits recorded at $t_k$, the
model's negative log-likelihood (NLL) is
\begin{equation}
	\mathcal L_{\mathrm{model}}(x,L)
	=-\sum_k n_k\log q_k.
	\label{eq:model-nll}
\end{equation}

Raw NLL values depend on both the number of observations and the entropy
of the observed exit-time distribution, both of which can vary with $L$.
To compare model agreement across interval widths, we subtract the NLL
of the empirical distribution and normalize by the number of observations.
For $N=\sum_k n_k$ observations, the empirical exit-time probabilities
$p_k=n_k/N$ minimize the NLL over all normalized exit-time distributions,
giving the baseline
\begin{equation}
	\mathcal L_{\mathrm{emp}}
	=-\sum_{k:p_k>0}n_k\log p_k.
	\label{eq:emp-nll}
\end{equation}
The resulting excess NLL per observation is the
Kullback--Leibler (KL) divergence from the observed to the predicted exit-time
distribution:\cite{KullbackLeibler1951}
\begin{equation}
	\begin{aligned}
		D_{\mathrm{KL}}^{\mathrm{exit}}(x,L)
		 & =\frac{\mathcal L_{\mathrm{model}}-\mathcal L_{\mathrm{emp}}}{N} \\
		 & =\sum_{k:p_k>0}p_k\log\frac{p_k}{q_k}.
	\end{aligned}
	\label{eq:excess-nll}
\end{equation}

The KL divergence is zero when the observed and predicted distributions
agree; larger values indicate poorer agreement. Disagreement can arise
from variation of diffusivity within the interval, a residual gradient
of the effective PMF, or memory effects beyond the Markovian
approximation. Sampling fluctuations also contribute to the measured
divergence.\cite{Paninski2003InformationEstimation} The score quantifies their
combined effect on the exit-time distribution without distinguishing
these causes.

\subsection{Global model agreement}

The exit-time KL divergence assesses the diffusion model within individual
residence intervals of the biased trajectory. To test whether the inferred
diffusivities also describe motion across these intervals, we compare
model propagators with those from independent, unbiased MD
trajectories.\cite{ChipotComer2016}
Solving the Smoluchowski equation for this comparison requires a
diffusivity profile throughout the sampled region. We therefore
interpolate the local estimates $\widehat D(x;L)$ at the selected interval
width $L$ to obtain a continuous, positive profile $D(x)$.

A propagator $p(x,t\mid x_0)$ is the conditional probability density of
finding a molecule at position $x$ after time $t$, given that it starts
at $x_0$. We obtain the model propagator by solving
\eqnref{eq:smoluchowski} with the inferred diffusivity profile $D(x)$ and
the unbiased PMF $F(x)$.
For a given starting position $x_0$ and lag time $t$, we quantify model
agreement using the KL divergence from the MD propagator to the model
propagator:
\begin{equation}
	D_{\mathrm{KL}}^{\mathrm{prop}}(t;x_0)
	=\int p_{\mathrm{MD}}\log\frac{p_{\mathrm{MD}}}{p_{\mathrm{model}}}
	\,\mathrm dx,
	\label{eq:propagator-kl}
\end{equation}
where $p_{\mathrm{MD}}$ and $p_{\mathrm{model}}$ denote the normalized
conditional position densities from unbiased MD and the diffusion model,
respectively, evaluated at the same $t$ and nominal starting position
$x_0$.

Like the exit-time score, this divergence measures excess NLL per
observation, making it a natural choice consistent
with likelihood-based analyses of diffusion
models.\cite{Hummer2005,GhyselsVenable2017} It is not the only measure of
model mismatch; we explore additional measures in
\secref{si-sec:propagation-metrics}.

Agreement over a range of lag times supports the use of the inferred
diffusivity profile in a Smoluchowski description of unbiased motion
within the sampled regions and time scales. This comparison also tests
whether diffusivities obtained from biased trajectories describe unbiased
dynamics.

Disagreement can reflect errors in the PMF or diffusivity
profile, bias-induced changes in molecular dynamics, or
memory effects beyond the Markovian approximation.\cite{Daldrop2017ExternalPotential,HarrisChipotRoux2024}
Sampling fluctuations also contribute to the measured propagator divergence.

\section{Computational Details}

\label{sec:computational-details}

We describe the membrane systems and trajectories first, then the
extraction of diffusivities and exit-time scores. The final two
subsections explain how the resulting profiles are used to calculate
propagators for independent validation and steady-state permeability.

\subsection{Molecular simulations}

We examine two membrane models with contrasting lipid organization. The
first is a bilayer of 1-palmitoyl-2-oleoyl-sn-glycero-3-phosphocholine
(POPC), a single-component model membrane in the fluid,
liquid-crystalline phase under our simulation
conditions.\cite{KucerkaTristramNagle2005}

\begin{figure*}[!tb]
	\centering
	\includegraphics[width=4.8in]{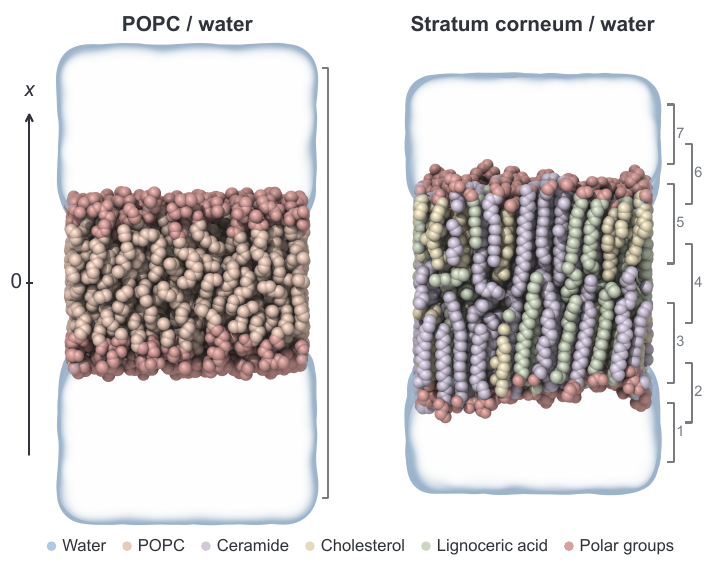}
	\caption{Snapshots of the hydrated POPC and SC membranes at the same
		spatial scale. The arrow marks the membrane normal $x$, with the
		origin at the membrane center of mass. Brackets mark the POPC
		sampling region and seven overlapping SC windows. Bounds are given
		in \tabref{si-tab:popc-abf} and \tabref{si-tab:sc-windows},
		respectively; SC window numbers match those shown here.}
	\label{fig:system-snapshots}
\end{figure*}

The second is a model of the skin's
stratum corneum (SC) lipid barrier based on the membrane model of Wang
and Klauda.\cite{WangKlauda2018SC} Its ternary mixture
of ceramide NS (24:0), cholesterol, and lignoceric acid represents the
short-periodicity phase of the SC lipid matrix. This membrane has highly
ordered lipid chains, low lipid mobility, and long-lived structural
asymmetries.\cite{Thomas2025Skin} \figref{fig:system-snapshots} shows the
chain organization and packing of both membranes.

We examine water as a permeant in both membranes. In the SC membrane,
we also examine
acetone and 6-methyl-5-hepten-2-one (6-MHO), two products of skin-oil
ozonolysis.\cite{WisthalerWeschler2010} Uptake into skin competes with
release into indoor air,\cite{vonDomaros2020Multiscale}
so their transport through the SC lipid barrier is relevant to models of
indoor chemistry and exposure.\cite{Lakey2019SkinClothing,Shiraiwa2019MOCCIE}

For the SC interval-width scans, we reused previously published ABF
trajectories. For POPC, we generated an ABF trajectory
for a new system containing 100 lipids and 8330 water molecules. Its larger
aqueous regions allow us to assess bulk-water diffusivity convergence
as the interval width increases. The POPC PMF, autocorrelation-based
diffusivity profiles, and unbiased MD propagators used for comparison
were taken from the previously studied system containing 72 lipids and
2967 water molecules.\cite{Thomas2026SC,Thomas2026RTA}

All systems used the CHARMM36 lipid force
field\cite{KlaudaVenable2010} and TIP3P
water.\cite{JorgensenChandrasekhar1983} Acetone and 6-MHO were described
by the CHARMM General Force Field
(CGenFF).\cite{VanommeslaegheHatcher2010} System compositions and box
dimensions are listed in \tabref{si-tab:popc-composition} for the new POPC
system and \tabref{si-tab:system-compositions} for SC, and settings
for the new POPC simulations are given in \tabref{si-tab:popc-protocol}.

In both sampling and analysis, $x$ is the displacement of the permeant's
center of mass from the membrane's center of mass, projected onto the
membrane normal. ABF sampling used one region for POPC and seven
overlapping windows for SC, each bounded by harmonic walls
(\figref{fig:system-snapshots}).

\subsection{Residence-time analysis}

To obtain the profiles and local agreement scores used for width
selection, we evaluate residence intervals of width $L$ from
\qty{0.5}{\angstrom} to \qty{15}{\angstrom} using permeant positions
recorded every \qty{2}{\femto\second}. For each interval, we use only
sampling windows whose walls enclose the entire interval. Every frame
with the permeant inside the interval serves as a time origin, from which
we measure the first-exit time $T$ through either boundary. We exclude
time origins with no observed exit before the trajectory ends. We pool
exit-time observations from intervals centered at $x$ and $-x$
in the two leaflets to improve sampling.

From the sampled first-exit times, we calculate the mean
$\widehat\tau(x;L)$, the diffusivity $\widehat D(x;L)$ using
\eqnref{eq:rta}, and the exit-time KL divergence using
\eqnref{eq:excess-nll}. We account for correlations between observations
by block averaging to estimate the standard error of
$\widehat\tau$,\cite{FlyvbjergPetersen1989,Jonsson2018} then propagate
this uncertainty to $\widehat D$. Details of window pooling and uncertainty
estimation are given in \secref{si-sec:rta-uncertainty}.

\subsection{Propagator calculations}

For independent validation, we compare RTA diffusivities with estimates
from velocity and position
autocorrelation functions (VACF and PACF) measured in harmonically
restrained simulations.\cite{WoolfRoux1994,Hummer2005} For propagator and
permeability calculations, we fit the discrete estimates from
each method to obtain continuous, positive diffusivity profiles. The
interpolation procedure is described in \secref{si-sec:profile-fitting}.
We construct continuous PMF profiles using the same smoothing procedure.
We symmetrize all diffusivity profiles, PMFs, and MD propagators about
the membrane center.

For each system, we use the same PMF (\figref{si-fig:pmf-comparison})
with each diffusivity profile. We restrict the propagator comparison
to the starting position $x_0=0$, where differences between diffusivity
estimates from the various methods are
largest.\cite{Thomas2026RTA,Thomas2026SC}

We solve \eqnref{eq:smoluchowski} by
spatial discretization and matrix exponential propagation, starting from
$p(x,0)=\delta(x)$ at the membrane center. The unbiased MD trajectories
approximate this initial condition: configurations with $|x|<\qty{0.05}{\angstrom}$
were selected from restrained simulations, and the restraint was removed
before propagation.
Numerical details are given in
\secref{si-sec:propagator-calculation}.

\subsection{Permeability calculations}

To assess the transport consequences of the diffusivity profiles, we
calculate permeability from \eqnref{eq:isd} using the same diffusivity
and PMF profiles as in the propagator calculations,
with bulk water as the reference phase.
We integrate between the symmetric positions where the water density
reaches half its bulk value. \secref{si-sec:permeability} gives the bounds
and describes how diffusivity uncertainties affect permeability.

\section{Results and Discussion}

\label{sec:results}

We first examine how the interval width affects the diffusivity profiles
and permeability. These results motivate a selection criterion that
combines bulk convergence with exit-time agreement. We develop the
interval-width selection criterion using POPC/water, apply the same
criterion without modification to the SC systems, and test the resulting
diffusivity profiles against independent, unbiased molecular trajectories.

\subsection{Interval-width dependence}

\begin{figure}[!htbp]
	\centering
	\includegraphics[width=3.0in]{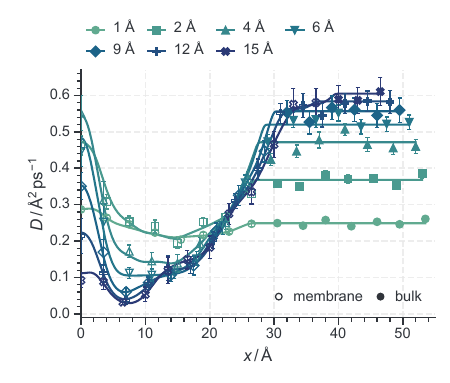}
	\caption{RTA diffusivity profiles for water across a POPC bilayer at
		different interval widths. Markers show RTA diffusivities; error bars
		indicate \qty{95}{\percent} confidence intervals. Lines show fitted profiles.}
	\label{fig:popc-water-rta-profiles}
\end{figure}

For water in the POPC system, the RTA diffusivity profile depends strongly
on the residence-interval width $L$
(\figref{fig:popc-water-rta-profiles}). In bulk water, diffusivity increases
with $L$, although the increase becomes smaller at larger widths.
Within the membrane, the response depends on position, so changing $L$
also changes the profile shape. At the membrane center, diffusivity rises
from $L=\qty{1}{\angstrom}$ to $\qty{4}{\angstrom}$ and then falls as the interval widens further.

Wider intervals probe longer exit times, over which molecular memory
may have less influence, consistent with the bulk trend. Within the
membrane, they also span regions with different diffusivities, challenging
RTA's assumption of constant diffusivity within each interval. Selecting
$L$ therefore requires balancing spatial resolution against the time
scales on which motion is consistent with the Markovian approximation.
The profile changes reflect this tradeoff but do not by themselves
identify a suitable width.

The diffusivity profiles for water, acetone, and 6-MHO in the SC membrane
show qualitatively similar width dependence
(\figref{si-fig:sc-rta-profiles}). These changes also affect permeability
through the integrated resistance to permeation.

\subsection{Consequences for permeability}

\begin{figure}[!htbp]
	\centering
	\includegraphics[width=3.0in]{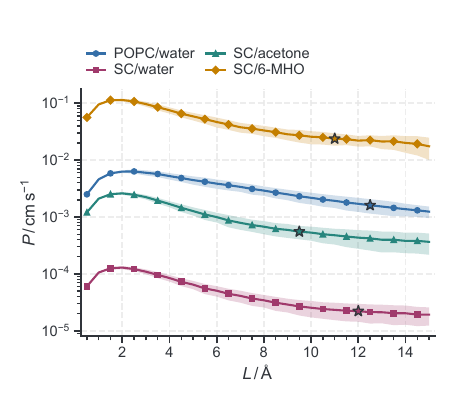}
	\caption{RTA permeability in the four membrane/permeant systems, with
		the PMF held fixed for each system. Stars mark the widths selected
		by the criterion in \secref{sec:width-selection}. Shaded bands show
		sensitivity to pointwise diffusivity uncertainties
		(\secref{si-sec:permeability}).}
	\label{fig:permeability-convergence}
\end{figure}

Permeability depends strongly on interval width in all four systems
(\figref{fig:permeability-convergence}). With the PMF and integration
bounds fixed for each system, this dependence arises entirely from
the diffusivity profiles. Their effect on permeability is greatest in
regions that contribute substantially to the integrated ISD resistance
(\eqnref{eq:isd}).
For all four systems, permeability initially rises to a maximum at
$L=\qtyrange{2}{2.5}{\angstrom}$, then falls by approximately five- to
sevenfold by $L=\qty{15}{\angstrom}$. Interval-width selection is therefore
essential for interpreting the permeability obtained from RTA.

This dependence generally weakens at larger $L$, most noticeably for
SC/water, whereas POPC/water continues to decrease appreciably. A flatter
permeability curve, however, does not establish that the diffusion model
describes molecular motion more accurately, nor does a lower value of $P$.
Longer exit times can also reduce the number of effectively independent
observations in a finite trajectory, making sampling uncertainty relevant
at larger widths. We therefore base interval-width selection on
bulk-diffusivity convergence and exit-time agreement.

\subsection{Interval-width selection}

\label{sec:width-selection}

The selection criterion uses bulk convergence to define the eligible
widths and local exit-time agreement to choose among them. We formulate
it for POPC/water and apply it without modification to the SC systems.
Bulk water provides a reference because its diffusivity should be
approximately independent of position and stable over a range of $L$.
We use permissive heuristic tolerances to identify widths consistent
with this behavior without excluding potentially suitable choices.

\begin{figure}[!htbp]
	\centering
	\includegraphics[width=3.0in]{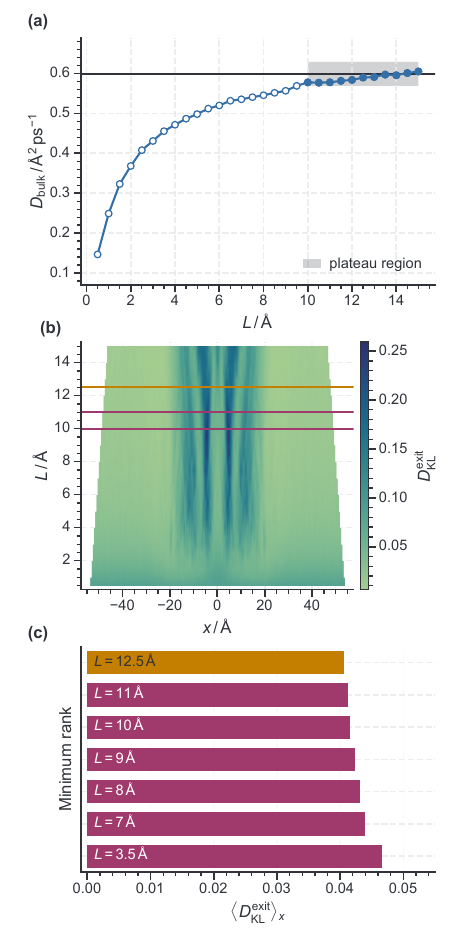}
	\caption{Interval-width diagnostics for POPC/water. (a) Bulk RTA
		diffusivity. Filled markers identify bulk-compatible widths; the
		line and shaded band show the representative bulk value and its
		$\pm\qty{5}{\percent}$ tolerance. (b) Exit-time KL divergence
		$D_{\mathrm{KL}}^{\mathrm{exit}}$. (c) Local minima of the spatial mean
		$\langle D_{\mathrm{KL}}^{\mathrm{exit}}\rangle_x$, ranked from lowest
		to highest. Horizontal lines in (b) mark the three lowest-scoring
		minima. Gold marks the selected interval width
		$L=\qty{12.5}{\angstrom}$, which also gives the global minimum.
		Spatial means use a fixed domain
		at every width; spatial averaging and gap treatment are described
		in \secref{si-sec:rta-spatial-average}.}
	\label{fig:popc-water-selection}
\end{figure}

For each $L$, we first locate a spatial diffusivity plateau, marked by
filled symbols in \figref{fig:popc-water-rta-profiles}. We then compare
the representative bulk values across widths
(\figpanelref{fig:popc-water-selection}{a}). For POPC/water, these values
satisfy the stability criteria over $L=\qtyrange{10}{15}{\angstrom}$,
which defines the bulk-compatible range. Details of plateau detection and
convergence assessment are given in \secref{si-sec:bulk-convergence}.

Bulk convergence alone does not establish that the same widths describe
motion within the membrane, where diffusivity varies with position.
For each interval, the RTA diffusivity reproduces the measured mean exit
time by construction, but this does not guarantee agreement with the full
exit-time distribution. We therefore assess local model agreement using
the exit-time KL divergence defined in \eqnref{eq:excess-nll}, shown as a
function of position and interval width in
\figpanelref{fig:popc-water-selection}{b}.
The map shows how the mismatch varies with position at each width.
We average over a common spatial domain to compare widths,
obtaining $\langle D_{\mathrm{KL}}^{\mathrm{exit}}\rangle_x$.
Local minima identify widths with better agreement than their neighbors;
\figpanelref{fig:popc-water-selection}{c} ranks these minima.

The combined criterion selects the interval width at the local minimum of
$\langle D_{\mathrm{KL}}^{\mathrm{exit}}\rangle_x$ with the lowest value
within the bulk-compatible range. We refer to the width chosen by this
criterion as the selected interval width. For POPC/water, this selects
$L=\qty{12.5}{\angstrom}$, which also gives the global minimum across
all tested widths. Thus, the width with the best spatially averaged
exit-time agreement also satisfies bulk convergence.

\begin{figure}[!htbp]
	\centering
	\includegraphics[width=3.0in]{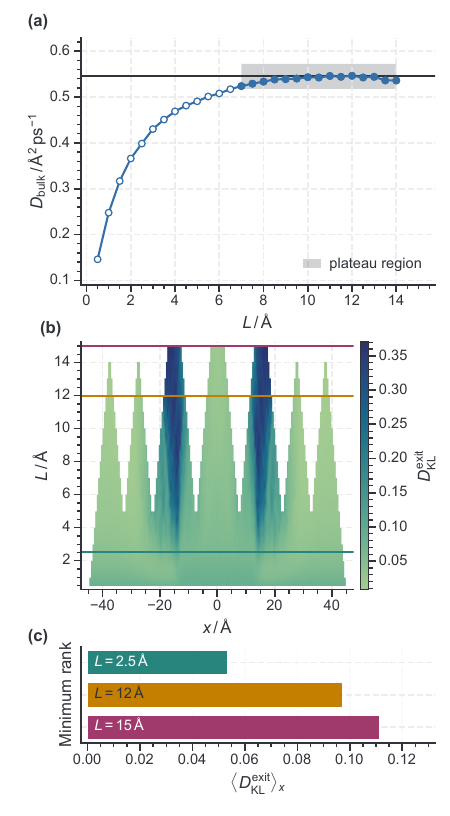}
	\caption{Interval-width diagnostics for SC/water. Panels and shared
		encodings follow \figref{fig:popc-water-selection}. Gold marks the selected
		interval width $L=\qty{12}{\angstrom}$; teal marks the global minimum
		at $L=\qty{2.5}{\angstrom}$, outside the bulk-compatible range.
		Gaps in (b) mark residence intervals not fully contained within
		any sampling window.}
	\label{fig:sc-water-selection}
\end{figure}

The SC membrane provides a more challenging environment because of its
structural complexity and slow lipid dynamics.\cite{Thomas2025Skin}
For SC/water, the global minimum of
$\langle D_{\mathrm{KL}}^{\mathrm{exit}}\rangle_x$
occurs at $L=\qty{2.5}{\angstrom}$
(\figpanelref{fig:sc-water-selection}{b–c}), outside the bulk-compatible
range in \figpanelref{fig:sc-water-selection}{a}. Requiring bulk convergence
therefore changes the selection: the combined criterion selects
$L=\qty{12}{\angstrom}$, the width at the local minimum with the lowest
spatial mean within the bulk-compatible range.

Acetone and 6-MHO in the SC membrane show the same distinction between
the width at the global exit-time KL minimum and the selected interval width
(\figrefs{si-fig:sc-acetone-selection}{si-fig:sc-6mho-selection}).
Their global minima occur at the same narrow width as for SC/water,
outside their respective bulk-compatible ranges. The combined criterion
instead selects wider intervals for both permeants.
\tabref{tab:interval-width-selection} summarizes the bulk-compatible
ranges, widths at the global minima, and selected interval widths for all
four systems.
Because interval-width selection uses only biased trajectories, testing
whether the resulting profiles predict unbiased motion provides an
independent assessment of the criterion.

\begin{table}[!htbp]
	\centering
	\caption{Interval-width selection. Minima refer to
		$\langle D_{\mathrm{KL}}^{\mathrm{exit}}\rangle_x$. The selected width
		is the width at the lowest local minimum within the bulk-compatible
		range. All entries are interval widths $L$ in \unit{\angstrom}.}
	\label{tab:interval-width-selection}
	\small
	\setlength{\tabcolsep}{4pt}
	\begin{tabular}{lrrr}
		\toprule
		System     & Bulk range           & \shortstack{Width at\\global\\minimum} & \shortstack{Selected\\width} \\
		\midrule
		POPC/water & \numrange{10}{15}    & \num{12.5}                             & \num{12.5}                   \\
		SC/water   & \numrange{7}{14}     & \num{2.5}                              & \num{12.0}                   \\
		SC/acetone & \numrange{5.5}{10.5} & \num{2.5}                              & \num{9.5}                    \\
		SC/6-MHO   & \numrange{5}{11}     & \num{2.5}                              & \num{11.0}                   \\
		\bottomrule
	\end{tabular}
\end{table}

\subsection{Independent propagator validation}

\begin{figure*}[!tb]
	\centering
	\makebox[\textwidth][c]{\includegraphics[width=7in]{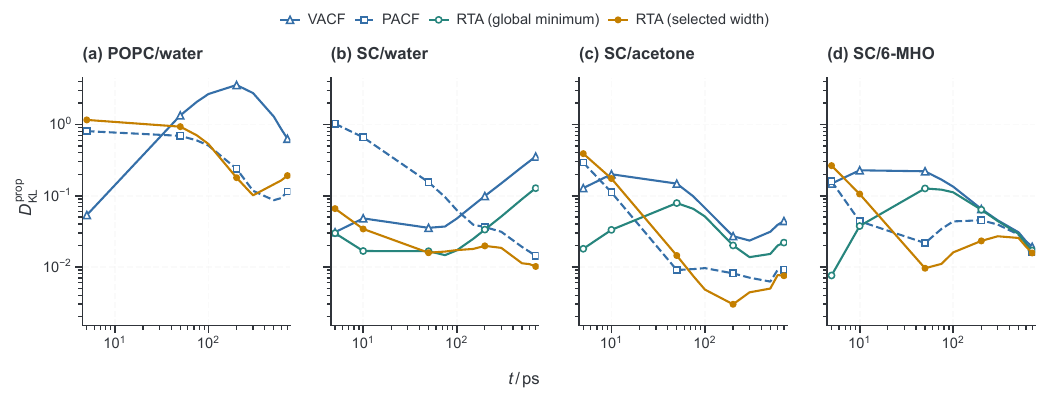}}
	\caption{Propagator KL divergence from unbiased MD for (a) POPC/water,
		(b) SC/water, (c) SC/acetone, and (d) SC/6-MHO. Gold denotes RTA at
		the selected interval widths (\tabref{tab:interval-width-selection}); teal
		denotes RTA at the width of the global minimum of
		$\langle D_{\mathrm{KL}}^{\mathrm{exit}}\rangle_x$,
		$L=\qty{2.5}{\angstrom}$, for the SC systems.
		For POPC/water, the selected width also gives this global minimum.}
	\label{fig:propagator-kl}
\end{figure*}

\begin{figure*}[!tb]
	\centering
	\makebox[\textwidth][c]{\includegraphics[width=7in]{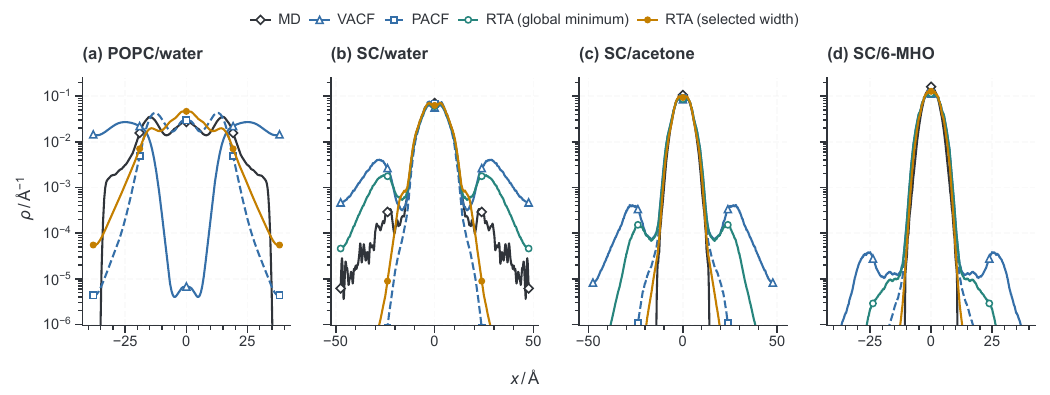}}
	\caption{Unbiased MD and model propagators at
		$t=\qty{200}{\pico\second}$, starting at $x_0=0$. Systems and method
		encodings follow \figref{fig:propagator-kl}.}
	\label{fig:propagator-density-examples}
\end{figure*}

We compare model propagators with independent, unbiased MD trajectories
starting at the membrane center over lag times of
\qtyrange{50}{700}{\pico\second}. This tests predictions beyond the local
exit-time statistics used for width selection. Alongside the RTA profiles
at the selected interval widths, we include previously reported VACF and
PACF profiles,\cite{Thomas2026RTA,Thomas2026SC} keeping the PMF fixed within
each system. For SC, we also include RTA profiles at the width of the
global exit-time KL minimum to assess the effect of requiring bulk
convergence.
All profiles are shown in \secref{si-sec:diffusivity-comparison}.
\figref{fig:propagator-kl} compares the resulting propagator KL
divergences from MD, $D_{\mathrm{KL}}^{\mathrm{prop}}(t;0)$, defined in
\eqnref{eq:propagator-kl}, with lower values indicating closer agreement.

Requiring bulk convergence improves longer-time predictions in all three
SC systems compared with using the width at the global exit-time KL
minimum. For acetone and 6-MHO, the RTA profiles at the selected widths
give lower propagator KL divergence throughout the tested range
(\figpanelref{fig:propagator-kl}{c--d}). For water, the two RTA models
give nearly equal divergences at \qty{100}{\pico\second}, but the model
using the selected width agrees more closely with MD at every longer lag
time (\figpanelref{fig:propagator-kl}{b}). The global exit-time KL minimum
therefore does not by itself identify the profile that best predicts
unbiased motion at longer lag times.

RTA at the selected interval widths is also competitive with the
autocorrelation-based methods.
It gives closer agreement with MD than VACF in all four systems at every
tested lag time. RTA also generally outperforms PACF in the SC systems:
the only exception is acetone at the shortest tested lag time
(\figpanelref{fig:propagator-kl}{b--d}). In POPC/water, RTA and PACF give
comparable agreement, with their ranking changing across lag times
(\figpanelref{fig:propagator-kl}{a}). Their ordering also depends on the
discrepancy metric, with several alternative metrics favoring PACF
(\secref{si-sec:propagation-metrics}). These observations are consistent
with our earlier RTA study, which finds that the methods employed here
do not yield a single lag-time-independent diffusion model that
reproduces POPC/water dynamics across all tested time
scales.\cite{Thomas2026RTA}

The propagators in \figref{fig:propagator-density-examples} show how
the inferred diffusivities affect the distribution of probability between
the membrane center and surrounding regions. For POPC/water,
VACF predicts too little probability density at the center and too much
toward the aqueous regions, indicating excessive spreading. RTA at the
selected interval width and PACF retain the central peak but underestimate
the MD tails
(\figpanelref{fig:propagator-density-examples}{a}). The remaining
discrepancies are larger in POPC/water
than in SC, consistent with the KL comparisons in \figref{fig:propagator-kl}.

For SC, the main benefit of requiring bulk convergence is a reduction
in excessive spreading. At the width of the global exit-time KL minimum,
the propagators assign too much probability away from the membrane center.
For water, this appears as pronounced shoulders toward the aqueous regions. The
RTA profile at the selected width reduces these shoulders, although its
propagator still underestimates the far tails
(\figpanelref{fig:propagator-density-examples}{b}). For acetone and
6-MHO, the RTA profiles at the selected widths reduce the excessive tails
while retaining similar agreement near the center
(\figpanelref{fig:propagator-density-examples}{c--d}). Because the PMF is
fixed within each comparison, the improved agreement reflects how the
diffusivity profiles at the selected widths describe molecular spreading
through the membrane.

The comparisons assess symmetrized motion from $x_0=0$ over
\qtyrange{50}{700}{\pico\second}. An earlier bilayer-center displacement test
examined motion over
\qty{5}{\pico\second}.\cite{GaalswykAwoonorWilliams2016}
At the longer lag times considered here, permeants spread beyond the
central residence interval and encounter regions of different mobility.
The propagators therefore
probe the combined effect of the spatially varying $D(x)$ profile over
the regions reached from the membrane center.

Agreement with these independent trajectories also supports the dynamics
underlying the RTA permeability estimates at the selected interval widths.
Within the ISD description, transient spreading and steady-state
permeation depend on the same PMF and diffusivity
profile.\cite{HarrisChipotRoux2024} Although
propagator agreement does not directly establish steady-state
permeability, it provides an independent test of this shared transport
description. The combined selection criterion thus yields diffusivity
profiles that predict unbiased motion beyond the local exit-time
statistics used to select them.

\section{Conclusions}

Combining bulk-diffusivity convergence with local exit-time agreement
provides a practical criterion for choosing the residence-interval width
in RTA. Applied without modification to the POPC and SC systems, it
identifies diffusivity profiles that predict unbiased motion beyond the
local exit statistics used for selection. In SC, requiring bulk
convergence improves longer-time propagator agreement over using the width
at the global exit-time KL minimum. RTA at the selected interval widths
gives closer agreement than VACF in all four systems and generally
outperforms PACF in SC; POPC rankings depend on lag time and discrepancy
metric. This is
consistent with our earlier finding that these methods do not yield a
single lag-time-independent diffusion model for POPC/water across all
tested time scales.\cite{Thomas2026RTA} Together with RTA's simple
estimation procedure, these results make interval-width selection a
useful step toward reliable position-dependent diffusivity profiles and
the permeability predictions derived from them.

An intriguing observation emerges in the SC systems: water, acetone, and
6-MHO all share a global exit-time KL minimum at
$L=\qty{2.5}{\angstrom}$. This common width lies close to the permeability
maxima at $L=\qty{2}{\angstrom}$, yet the corresponding diffusivity
profiles predict excessive spreading at longer times. Thus, chemically
different permeants show a similar preferred width for local exit-time
agreement, even though wider intervals give better predictions of
motion through the membrane.

We conjecture that this pattern reflects a separation between rapid
local motion and slower displacement between molecular environments.
Narrow intervals may capture diffusion-like motion within a local
environment, whereas wider intervals also probe displacement involving
environmental rearrangements. This raises a concrete follow-up question:
does the shared minimum track a characteristic length scale of the SC
lipid environment as membrane packing and mobility change? Comparing
width-dependent exit statistics with independent measurements of local
structure and environmental relaxation would test this interpretation.
The present results motivate that test but do not establish its
mechanism: exit-time mismatch can also arise from spatial heterogeneity,
residual bias, or sampling uncertainty.

A further opportunity is to test the criterion in other inhomogeneous
systems with a suitable bulk reference. Independent propagator
comparisons can establish when bulk convergence and exit-time agreement
provide sufficient guidance for diffusivity estimation in new dynamical
environments. With broader validation, this combination could offer a
routine way to select profiles for transport calculations, reducing the
need to repeat demanding propagation tests for every analysis.

\begin{suppinfo}
	Simulation details, analysis procedures, interval-width diagnostics,
	diffusivity and PMF profiles, propagator comparisons, and permeability
	and PMF sensitivity tests (PDF)
\end{suppinfo}

\begin{acknowledgement}
	The authors thank Douglas J. Tobias for helpful discussions.

	The authors used Codex to assist with manuscript editing, code development, and figure preparation. The authors reviewed and verified all text, code, and figures and bear sole responsibility for the entire content of this manuscript, including its accuracy, scientific interpretation, and conclusions.
\end{acknowledgement}

\section*{Data Availability}

All analyses are implemented in the PDDA Python library.\cite{PDDASoftware}
Simulation inputs, including coordinate files, topologies, force-field
parameters, and molecular dynamics run files, are provided in the following
repositories. Inputs for the previously published SC simulations reused in
this study are available at
\url{https://github.com/mvondomaros-lab/rta-paper-data}.
Inputs for the POPC simulations are available at
\url{https://github.com/mvondomaros-lab/interval-paper-data}.

\clearpage
\onecolumn
\newgeometry{total={6in,8.5in}}
\fontsize{11}{13.2}\selectfont
\pagenumbering{arabic}
\renewcommand{\thepage}{S\arabic{page}}
\setcounter{section}{0}
\setcounter{figure}{0}
\setcounter{table}{0}
\setcounter{equation}{0}
\renewcommand{\thesection}{S\arabic{section}}
\renewcommand{\thefigure}{S\arabic{figure}}
\renewcommand{\thetable}{S\arabic{table}}
\renewcommand{\theequation}{S\arabic{equation}}
\renewcommand{\theHsection}{si.\arabic{section}}
\renewcommand{\theHfigure}{si.\arabic{figure}}
\renewcommand{\theHtable}{si.\arabic{table}}
\renewcommand{\theHequation}{si.\arabic{equation}}
\AddToHook{cmd/section/before}{\Needspace{7\baselineskip}}
\AddToHook{cmd/subsection/before}{\Needspace{5\baselineskip}}
\begin{center}
{\Large\bfseries Supporting Information\par}
\medskip
{\large Choosing Spatial Resolution in the Residence-Time Approach to Local Diffusivity\par}
\end{center}
\section*{Contents}
\noindent\hyperref[si-sec:simulations]{\makebox[2.8em][l]{\ref*{si-sec:simulations}}POPC simulations}\nobreak\leaders\hbox{.}\hfill\pageref*{si-sec:simulations}\par
\noindent\hyperref[si-sec:rta-uncertainty]{\makebox[2.8em][l]{\ref*{si-sec:rta-uncertainty}}RTA uncertainty estimation}\nobreak\leaders\hbox{.}\hfill\pageref*{si-sec:rta-uncertainty}\par
\noindent\hyperref[si-sec:rta-implementation]{\makebox[2.8em][l]{\ref*{si-sec:rta-implementation}}Interval-width selection}\nobreak\leaders\hbox{.}\hfill\pageref*{si-sec:rta-implementation}\par
\noindent\hyperref[si-sec:sc-width-diagnostics]{\makebox[2.8em][l]{\ref*{si-sec:sc-width-diagnostics}}SC interval-width diagnostics}\nobreak\leaders\hbox{.}\hfill\pageref*{si-sec:sc-width-diagnostics}\par
\noindent\hyperref[si-sec:profile-fitting]{\makebox[2.8em][l]{\ref*{si-sec:profile-fitting}}Continuous diffusivity profiles}\nobreak\leaders\hbox{.}\hfill\pageref*{si-sec:profile-fitting}\par
\noindent\hyperref[si-sec:reference-pmfs]{\makebox[2.8em][l]{\ref*{si-sec:reference-pmfs}}Potentials of mean force}\nobreak\leaders\hbox{.}\hfill\pageref*{si-sec:reference-pmfs}\par
\noindent\hyperref[si-sec:propagator-calculation]{\makebox[2.8em][l]{\ref*{si-sec:propagator-calculation}}Numerical propagator calculation}\nobreak\leaders\hbox{.}\hfill\pageref*{si-sec:propagator-calculation}\par
\noindent\hyperref[si-sec:diffusivity-comparison]{\makebox[2.8em][l]{\ref*{si-sec:diffusivity-comparison}}Diffusivity profile comparison}\nobreak\leaders\hbox{.}\hfill\pageref*{si-sec:diffusivity-comparison}\par
\noindent\hyperref[si-sec:propagation-metrics]{\makebox[2.8em][l]{\ref*{si-sec:propagation-metrics}}Additional propagator discrepancy metrics}\nobreak\leaders\hbox{.}\hfill\pageref*{si-sec:propagation-metrics}\par
\noindent\hyperref[si-sec:permeability]{\makebox[2.8em][l]{\ref*{si-sec:permeability}}Permeability sensitivity to diffusivity uncertainty}\nobreak\leaders\hbox{.}\hfill\pageref*{si-sec:permeability}\par
\noindent\hyperref[si-sec:popc-pmf-sensitivity]{\makebox[2.8em][l]{\ref*{si-sec:popc-pmf-sensitivity}}POPC PMF sensitivity}\nobreak\leaders\hbox{.}\hfill\pageref*{si-sec:popc-pmf-sensitivity}\par
\noindent\hyperref[si-sec:software]{\makebox[2.8em][l]{\ref*{si-sec:software}}Software}\nobreak\leaders\hbox{.}\hfill\pageref*{si-sec:software}\par
\clearpage
\section{POPC simulations}

\label{si-sec:simulations}

\subsection{Simulation protocol}

\label{si-sec:simulation-setup}

\begin{table}[htbp]
  \centering
  \caption{Composition and simulation box dimensions of the new
    100-POPC/water system. Box lengths are given along the membrane
    normal $x$ and lateral directions $y$ and $z$. The tagged water
    permeant is included
    in the total water count.}
  \label{si-tab:popc-composition}
  \small
  \setlength{\tabcolsep}{5pt}
  \begin{tabular}{lrrlrrr}
    \toprule
    System     & Lipids    & Water      & Permeant       & $L_x$ / \unit{\angstrom} & $L_y$ / \unit{\angstrom} & $L_z$ / \unit{\angstrom} \\
    \midrule
    POPC/water & \num{100} & \num{8330} & 1 tagged water & \num{109.23}             & \num{58.44}              & \num{58.45}              \\
    \bottomrule
  \end{tabular}
\end{table}

\begin{table}[H]
  \centering
  \caption{System preparation and molecular dynamics (MD) settings for
    the new 100-POPC/water system.}
  \label{si-tab:popc-protocol}
  \small
  \begin{tabular}{@{}p{2.05in}p{3.65in}@{}}
    \toprule
    Setting                       & Value                                                                                                                        \\
    \midrule
    \multicolumn{2}{@{}l}{\textbf{System preparation}}                                                                                                           \\
    Construction                  & CHARMM-GUI Membrane Builder\cite{Jo2009MembraneBuilder}                                                                      \\
    Force field and water model   & CHARMM36 lipids;\cite{KlaudaVenable2010} TIP3P water\cite{JorgensenChandrasekhar1983}                                        \\
    Energy minimization           & \num{50000} steepest-descent steps                                                                                           \\
    Equilibration                 & \qty{20}{\nano\second} at constant pressure, followed by \qty{100}{\nano\second} at constant volume                          \\
    \midrule
    \multicolumn{2}{@{}l}{\textbf{MD settings}}                                                                                                                  \\
    Temperature                   & \qty{308}{\kelvin}                                                                                                           \\
    Pressure during equilibration & \qty{1}{\bar}; Langevin piston\cite{FellerZhang1995Piston} with relaxation time \qty{100}{\pico\second}                      \\
    Thermostat                    & Stochastic velocity rescaling;\cite{BussiDonadio2007} coupling time \qty{1}{\pico\second}                                    \\
    Integration                   & r-RESPA;\cite{TuckermanBerne1992} \qty{2}{\femto\second} time step                                                           \\
    Force evaluation intervals    & \qty{2}{\femto\second} for short-range forces; \qty{4}{\femto\second} for full electrostatics                                \\
    Bond constraints              & SETTLE for water;\cite{MiyamotoKollman1992SETTLE} SHAKE-H for other bonds involving hydrogen\cite{RyckaertCiccotti1977SHAKE} \\
    Van der Waals interactions    & Force switching over \qtyrange{10}{12}{\angstrom}; cutoff \qty{12}{\angstrom}                                                \\
    Neighbor-list radius          & \qty{16}{\angstrom}                                                                                                          \\
    Electrostatics                & Particle-mesh Ewald (PME);\cite{Essmann1995PME} interpolation order 6; maximum grid spacing \qty{1}{\angstrom}               \\
    \bottomrule
  \end{tabular}
\end{table}

\FloatBarrier

\subsection{Adaptive biasing force sampling}

\label{si-sec:abf-convergence}

\begin{table}[H]
  \centering
  \caption{Adaptive biasing force (ABF) production and trajectory-analysis
    settings for the 100-POPC/water system. ABF follows the average-force
    approach.\cite{Darve2001ABF,ComerGumbart2015} The final trajectory
    segment supplies the residence-time approach (RTA) estimates.}
  \label{si-tab:popc-abf}
  \small
  \begin{tabular}{@{}p{2.05in}p{3.65in}@{}}
    \toprule
    Setting                    & Value                                                                                                    \\
    \midrule
    Ensemble                   & NVT; box dimensions in \tabref{si-tab:popc-composition}                                                     \\
    Total ABF duration         & \qty{4000}{\nano\second}                                                                                 \\
    Sampling region            & $-\qty{54.1}{\angstrom} \leq x \leq \qty{54.1}{\angstrom}$                                               \\
    Bin width                  & \qty{0.1}{\angstrom}                                                                                     \\
    Bias activation            & Full bias after 1000 force samples per bin                                                               \\
    Bias updates               & Throughout production sampling                                                                           \\
    Boundary restraints        & Harmonic walls at the sampling boundaries; force constant \qty{500}{\kcal\per\mole\per\angstrom\squared} \\
    \midrule
    Coordinate output interval & \qty{2}{\femto\second} for $x$                                                                           \\
    RTA segment                & \qtyrange{3000}{4000}{\nano\second} (\qty{1000}{\nano\second} analyzed)                                  \\
    \bottomrule
  \end{tabular}
\end{table}

\begin{figure}[htbp]
  \centering
  \includegraphics[width=3.25in]{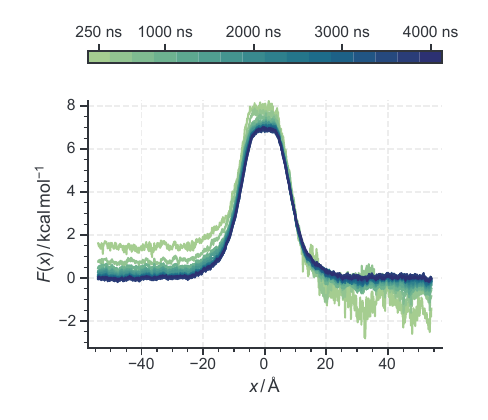}
  \caption{Convergence of the potential of mean force (PMF) during ABF
    sampling of the 100-POPC/water
    system. Each color-bar block represents one profile calculated from
    all data collected up to the indicated time. Profiles share a reference
    defined by their mean in the bulk-water regions $|x| \geq \qty{40}{\angstrom}$.
    The late-time profiles lie close together, supporting use of the
    final \qty{1000}{\nano\second} for RTA with a sufficiently converged ABF bias.}
  \label{si-fig:popc-abf-convergence}
\end{figure}

\FloatBarrier

\section{RTA uncertainty estimation}

\label{si-sec:rta-uncertainty}

The automated blocking procedure of Jonsson, based on
Flyvbjerg--Petersen blocking,\cite{FlyvbjergPetersen1989,Jonsson2018}
gives the standard error $\operatorname{SE}(\widehat\tau)$ while accounting
for correlations between first-exit observations that share an eventual
exit. At fixed interval width, first-order uncertainty propagation gives
\begin{equation}
  \frac{\operatorname{SE}(\widehat D)}{\widehat D}
  \simeq\frac{\operatorname{SE}(\widehat\tau)}{\widehat\tau},
  \label{si-eq:rta-uncertainty}
\end{equation}
For SC intervals sampled in multiple windows, we pool the observed
first-exit times, so each window contributes in proportion to its number
of observed exits. Uncertainties are combined for independent windows.
All RTA error bars denote approximate pointwise \qty{95}{\percent} confidence intervals,
$\widehat D\pm1.96\,\operatorname{SE}(\widehat D)$. They describe sampling
uncertainty at fixed $L$ and do not quantify departures from the local
diffusion model.

\FloatBarrier

\section{Interval-width selection}

\label{si-sec:rta-implementation}

\subsection{Spatial averaging}

\label{si-sec:rta-spatial-average}

The interval-width and interval-center grids both have
\qty{0.5}{\angstrom} spacing. Gaps in the exit-time KL maps mark
intervals that do not fit within any sampling window.

\begin{figure}[htbp]
  \centering
  \includegraphics[width=5.5in]{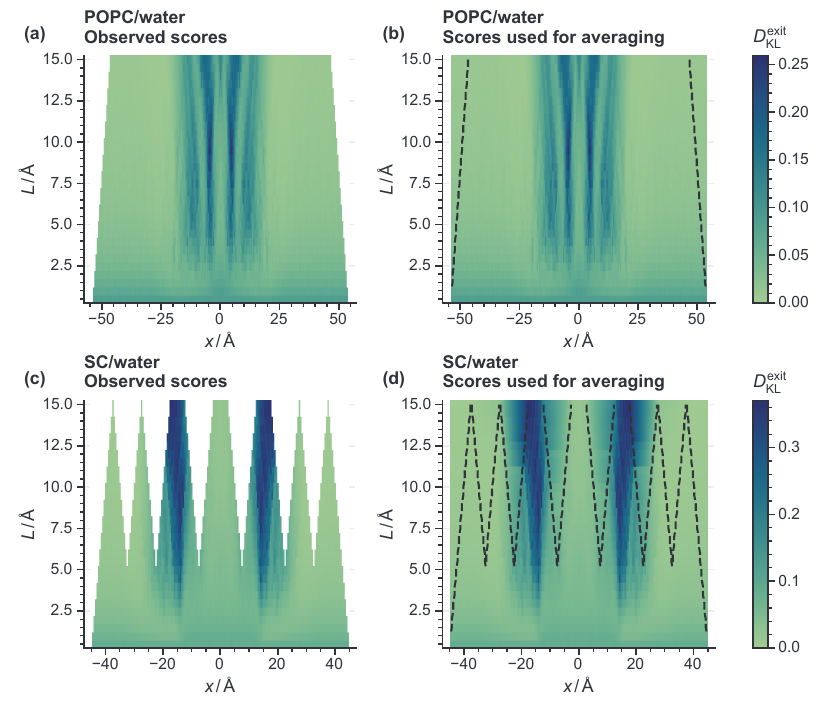}
  \caption{Exit-time KL maps for POPC/water (a, b) and SC/water (c, d),
    before (a, c) and after (b, d) gap filling for spatial averaging.
    Dashed lines outline the filled regions.}
  \label{si-fig:nll-gap-completion}
\end{figure}

For each width, we fill outer gaps by extending the nearest available
score and internal gaps by linear interpolation. Observed scores remain
unchanged. We then calculate $\langle D_{\mathrm{KL}}^{\mathrm{exit}}\rangle_x$
on the original grid of interval centers, with equal weight per center.
This fixes the averaging domain across widths, avoiding changes in the
spatial mean caused solely by loss of coverage. Gap filling is used only
for this score average; it does not supply diffusivity values for the
bulk-convergence diagnostic.
\figref{si-fig:nll-gap-completion} illustrates this procedure for POPC/water
and SC/water.

\FloatBarrier

\subsection{Bulk-diffusivity convergence}

\label{si-sec:bulk-convergence}

\tabref{si-tab:bulk-convergence-criteria} lists the bulk-convergence criteria
for the discrete diffusivity estimates.

A bulk plateau must lie in the outer half of the profile, reach the
outermost sampled position, and contain no gaps. Its overall trend and
typical scatter must each remain within a \qty{5}{\percent} tolerance. We retain the
widest qualifying plateau and calculate its bulk diffusivity by taking
the median of $\log D$ over the positions in that plateau, then
exponentiating. The central value of a width sequence is calculated
in the same way from its per-width bulk diffusivities. These median-based
statistics reduce the influence of outliers.\cite{RousseeuwCroux1993MAD}

Missing bulk values or insufficient plateau data break a width sequence.

\begin{table}[H]
  \centering
  \caption{Requirements for identifying a bulk-compatible interval-width
    range.}
  \label{si-tab:bulk-convergence-criteria}
  \small
  \begin{tabular}{@{}p{2.15in}p{3.50in}@{}}
    \toprule
    Criterion                         & Requirement                                                           \\
    \midrule
    Positions per plateau             & At least 3                                                            \\
    Consecutive interval widths       & At least 4                                                            \\
    Overall trend and typical scatter & Each at most \qty{5}{\percent}, within each plateau and across widths \\
    Deviation from the central value  & Every per-width bulk diffusivity within $\pm\qty{5}{\percent}$        \\
    \bottomrule
  \end{tabular}
\end{table}

If several sequences qualify, we choose the one with the highest central
bulk diffusivity. Ties favor the sequence containing more widths. We
extend this sequence to adjacent eligible widths whose bulk values
remain within \qty{5}{\percent} of the chosen value. The resulting bulk-compatible
range need not include the largest tested width. Within this range, we
choose the width at the local minimum of
$\langle D_{\mathrm{KL}}^{\mathrm{exit}}\rangle_x$ with the lowest value;
this is the selected interval width. Local exit-time KL
minima at either endpoint of the bulk-compatible range are eligible for
selection; the selected width can therefore coincide with a range boundary.

\FloatBarrier

\section{SC interval-width diagnostics}

\label{si-sec:sc-width-diagnostics}

\begin{table}[H]
  \centering
  \caption{Compositions and simulation box dimensions of the previously
    reported SC systems used for the RTA interval-width scans.\cite{Thomas2025Skin,Thomas2026RTA}
    Box lengths are given along the membrane normal $x$ and lateral
    directions $y$ and $z$. The tagged water permeant is included in the
    water count.}
  \label{si-tab:system-compositions}
  \small
  \setlength{\tabcolsep}{5pt}
  \begin{tabular}{lrrlrrr}
    \toprule
    System     & Lipids    & Water      & Permeant       & $L_x$ / \unit{\angstrom} & $L_y$ / \unit{\angstrom} & $L_z$ / \unit{\angstrom} \\
    \midrule
    SC/water   & \num{200} & \num{4800} & 1 tagged water & \num{95.14}              & \num{57.14}              & \num{57.14}              \\
    SC/acetone & \num{200} & \num{4800} & 1 acetone      & \num{95.35}              & \num{57.14}              & \num{57.14}              \\
    SC/6-MHO   & \num{200} & \num{4800} & 1 6-MHO        & \num{95.14}              & \num{57.14}              & \num{57.14}              \\
    \bottomrule
  \end{tabular}
\end{table}

\begin{table}[H]
  \centering
  \caption{SC ABF sampling windows. Bounds are measured
    along the membrane normal relative to the membrane center.}
  \label{si-tab:sc-windows}
  \small
  \setlength{\tabcolsep}{16pt}
  \begin{tabular}{crr}
    \toprule
    Window & $x_{\mathrm{min}}$ / \unit{\angstrom} & $x_{\mathrm{max}}$ / \unit{\angstrom} \\
    \midrule
    1      & \num{-45}                             & \num{-30}                             \\
    2      & \num{-35}                             & \num{-20}                             \\
    3      & \num{-25}                             & \num{-5}                              \\
    4      & \num{-10}                             & \num{10}                              \\
    5      & \num{5}                               & \num{25}                              \\
    6      & \num{20}                              & \num{35}                              \\
    7      & \num{30}                              & \num{45}                              \\
    \bottomrule
  \end{tabular}
\end{table}

For acetone and 6-MHO, the bulk-compatible ranges end before the largest
tested interval width. Increasing $L$ reduces the spatial support for
the bulk diagnostic: near the sampling walls, fewer interval centers
allow the entire residence interval to fit within a window. A qualifying
plateau can therefore lose the support required by
\tabref{si-tab:bulk-convergence-criteria}; an upper range boundary does
not by itself demonstrate loss of diffusive behavior.

\begin{figure}
  \centering
  \includegraphics[width=\linewidth]{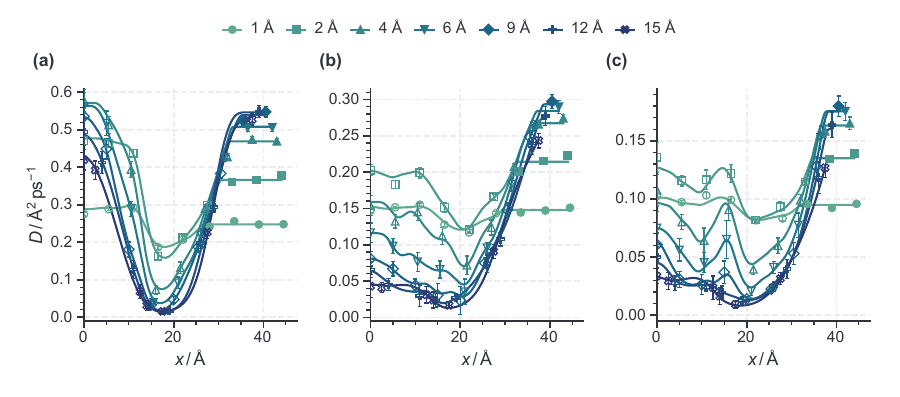}
  \caption{RTA diffusivity profiles for (a) water, (b) acetone, and
    (c) 6-MHO in the SC membrane. Markers show RTA diffusivities; error
    bars indicate \qty{95}{\percent} confidence intervals. Lines show fitted profiles
    constructed as described in \secref{si-sec:profile-fitting}.}
  \label{si-fig:sc-rta-profiles}
\end{figure}

\begin{figure}
  \centering
  \includegraphics[width=\linewidth]{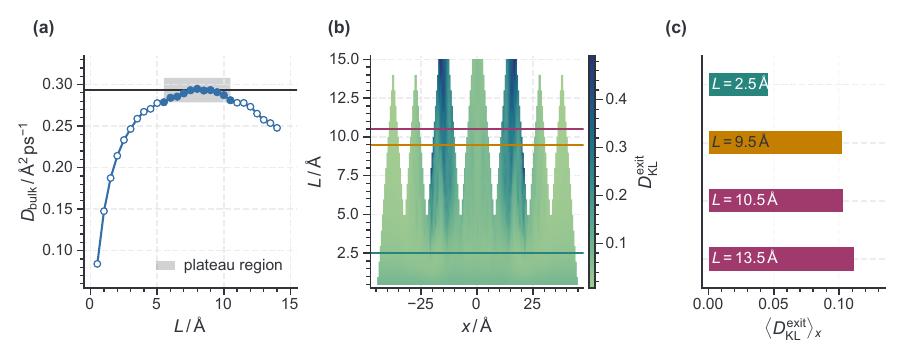}
  \caption{Interval-width diagnostics for SC/acetone. (a) Bulk
    diffusivity versus interval width. (b) Exit-time KL landscape.
    (c) Ranked local minima of
    $\langle D_{\mathrm{KL}}^{\mathrm{exit}}\rangle_x$.
    Gold marks the selected interval width $L=\qty{9.5}{\angstrom}$;
    teal marks the global minimum at $L=\qty{2.5}{\angstrom}$, outside
    the bulk-compatible range. Spatial averaging is specified in
    \secref{si-sec:rta-spatial-average}.}
  \label{si-fig:sc-acetone-selection}
\end{figure}

\begin{figure}
  \centering
  \includegraphics[width=\linewidth]{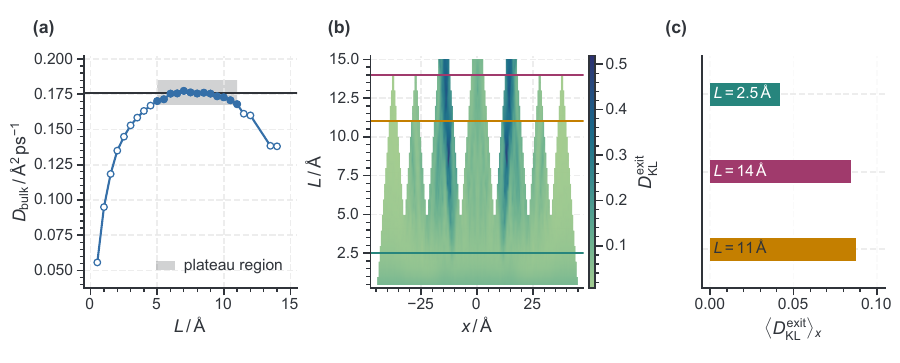}
  \caption{Interval-width diagnostics for SC/6-MHO. Panels follow
    \figref{si-fig:sc-acetone-selection}. Gold marks the
    selected interval width $L=\qty{11}{\angstrom}$; teal marks the
    global minimum at $L=\qty{2.5}{\angstrom}$, outside the
    bulk-compatible range.}
  \label{si-fig:sc-6mho-selection}
\end{figure}

\FloatBarrier

\section{Continuous diffusivity profiles}

\label{si-sec:profile-fitting}

Smoothing splines\cite{Wahba1990Splines} are fit to $\log D(x)$ with
an even-symmetry constraint and zero slope at the membrane center.
The smoothing length is \qty{1}{\angstrom} for RTA and
\qty{2}{\angstrom} for VACF and PACF.

To set the outer part of each fit, we search for a bulk plateau in the
outermost region with continuous data. We use the same \qty{5}{\percent} limits on
overall trend and typical scatter as in \secref{si-sec:bulk-convergence}.
If a plateau is found, the fitted profile joins its bulk value smoothly
and remains constant beyond the plateau onset. Both slope and curvature
are zero at the join. If no plateau is found, the profile reaches its
outer endpoint with zero slope and continues at that value.
This fitting step follows width selection and does not alter the discrete
estimates used in the selection diagnostics.

\FloatBarrier

\section{Potentials of mean force}

\label{si-sec:reference-pmfs}

\begin{figure}
  \centering
  \includegraphics[width=4.3in]{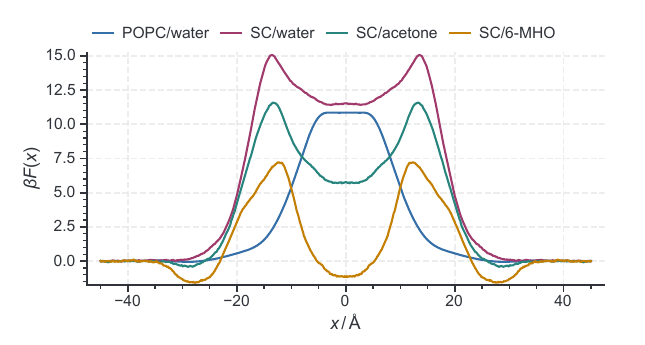}
  \caption{Reflection-symmetrized reduced potentials of mean force
    $\beta F(x)$ for POPC/water and the three SC systems.\cite{Thomas2026SC,Thomas2026RTA}
    Each profile is
    referenced to its bulk region.}
  \label{si-fig:pmf-comparison}
\end{figure}

\FloatBarrier

\begin{samepage}

  \section{Numerical propagator calculation}

  \label{si-sec:propagator-calculation}

  We discretize the Smoluchowski equation on a uniform periodic grid
  centered at $x=0$ and spanning the simulation box along $x$. An odd
  number of cells places one cell at the origin; the cell spacing $h$ is
  approximately \qty{0.1}{\angstrom}.
  For POPC/water, the propagation grid spans the 72-POPC box
  ($L_x=\qty{76.34}{\angstrom}$), matching the unbiased MD trajectories; the SC box lengths
  are given in \tabref{si-tab:system-compositions}. The transport calculations
  use \qty{310}{\kelvin} for POPC/water and \qty{305.15}{\kelvin} for the SC systems.
  Neighboring cells exchange probability through
  rates of the Wang--Peskin--Elston form:\cite{WangPeskinElston2003}
  \begin{equation}
    \begin{aligned}
      k_{i\to i+1} & = \frac{D_{i+1/2}}{h^2}B(u_{i+1}-u_i), \\
      k_{i+1\to i} & = \frac{D_{i+1/2}}{h^2}B(u_i-u_{i+1}),
    \end{aligned}
    \label{si-eq:discrete-rates}
  \end{equation}
\end{samepage}
where $u_i=\beta F(x_i)$, $B(v)=v/(e^v-1)$ with $B(0)=1$, and
$D_{i+1/2}=2D_iD_{i+1}/(D_i+D_{i+1})$ is the harmonic mean of
$D_i=D(x_i)$ and $D_{i+1}=D(x_{i+1})$. Indices wrap around at the periodic
boundary. The rates satisfy detailed balance with the discrete Boltzmann
distribution.

The matrix $\mathbf A$ has off-diagonal entries $A_{ji}=k_{i\to j}$
for neighboring cells; each diagonal entry is the negative sum of the
rates leaving that cell. The grid densities $\mathbf p(t)$ then obey
\begin{equation}
  \frac{d\mathbf p}{dt}=\mathbf A\mathbf p,
  \qquad
  \mathbf p(t)=e^{t\mathbf A}\mathbf p(0).
  \label{si-eq:discrete-propagation}
\end{equation}
This construction conserves probability, $h\sum_i p_i(t)=1$, and
preserves the stationary density $p_i^{\mathrm{eq}}\propto e^{-u_i}$.

We represent the delta initial condition by
$p_i(0)=1/h$ in the cell centered at $x=0$ and zero elsewhere.
At each requested lag time, SciPy's
\texttt{expm\_multiply}\cite{Virtanen2020SciPy} evaluates the matrix
exponential action using the Al-Mohy--Higham
algorithm.\cite{AlMohyHigham2011}

\FloatBarrier

\begin{samepage}

  \section{Diffusivity profile comparison}

  \label{si-sec:diffusivity-comparison}

  \begin{figure}[H]
    \centering
    \includegraphics[width=\linewidth]{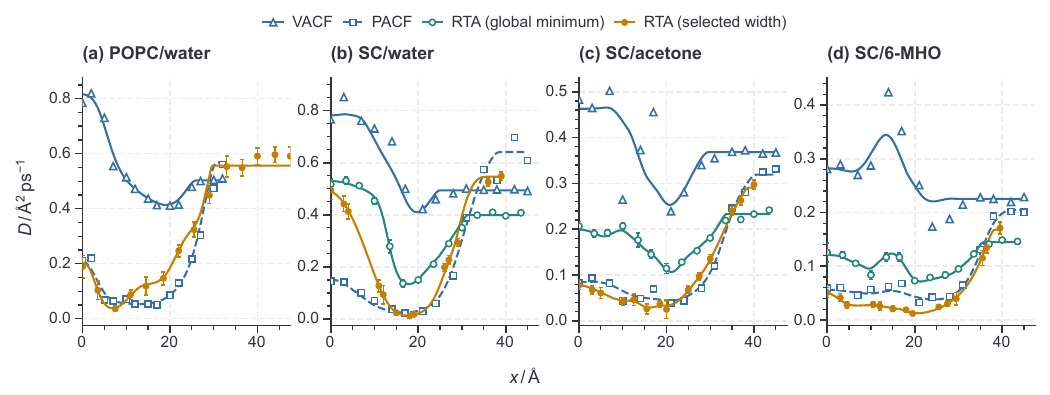}
    \caption{VACF, PACF, and RTA diffusivity profiles used for model
      propagation.
      Gold denotes RTA at the selected interval widths chosen by combining
      bulk convergence with exit-time KL agreement:
      $L=$\qtylist{12.5;12;9.5;11}{\angstrom} for
      POPC/water, SC/water, SC/acetone, and SC/6-MHO, respectively;
      teal denotes RTA at the width of the global minimum of
      $\langle D_{\mathrm{KL}}^{\mathrm{exit}}\rangle_x$,
      $L=\qty{2.5}{\angstrom}$, in each SC system.
      RTA error bars show \qty{95}{\percent} confidence intervals.}
    \label{si-fig:diffusivity-comparison}
  \end{figure}
\end{samepage}

\FloatBarrier

\section{Additional propagator discrepancy metrics}

\label{si-sec:propagation-metrics}

\begin{figure}[H]
  \centering
  \includegraphics[width=5.5in]{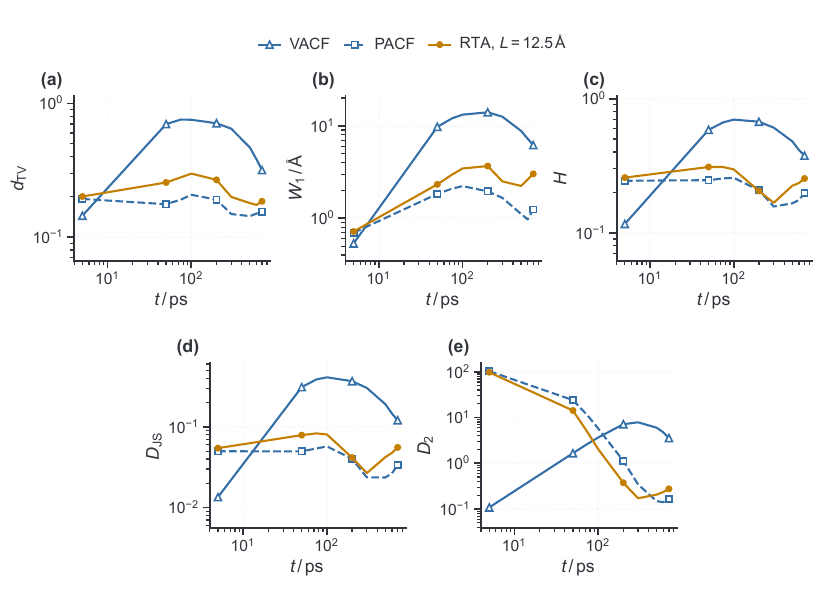}
  \caption{Additional comparisons of model propagators with unbiased MD
    for POPC/water, starting at the membrane center.}
  \label{si-fig:popc-water-propagation-metrics}
\end{figure}

The additional metrics are total variation ($d_{\mathrm{TV}}$),
Wasserstein-1 distance ($W_1$), Hellinger
distance ($H$),\cite{GibbsSu2002Metrics} Jensen--Shannon divergence
($D_{\mathrm{JS}}$),\cite{Lin1991JS} and Rényi divergence of order 2
($D_2$).\cite{VanErvenHarremoes2014}
All metrics vanish for identical distributions and increase with
disagreement. $W_1$ has units of length; the other metrics are dimensionless.
\figref{si-fig:popc-water-propagation-metrics} includes lag times from
\qty{5}{\pico\second} to \qty{700}{\pico\second}.

\begin{figure}[H]
  \centering
  \includegraphics[width=5.5in]{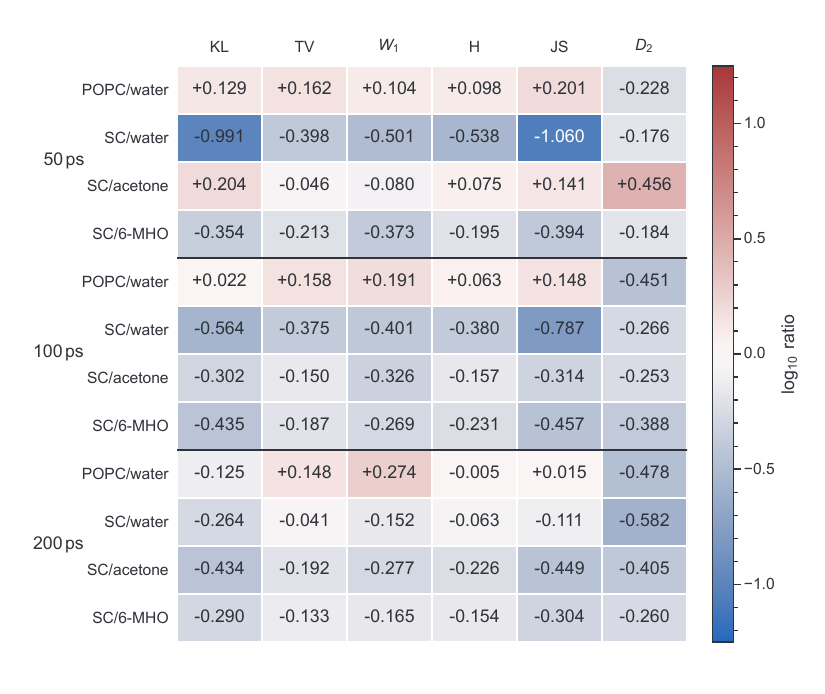}
  \caption{Propagation-metric comparison for RTA at the selected interval
    widths and PACF.
    Each cell gives $\log_{10}(d_{\mathrm{RTA}}/d_{\mathrm{PACF}})$, where
    $d$ is the discrepancy from the MD density under the indicated metric.
    Negative values favor RTA; positive values favor PACF. TV, $W_1$,
    H, JS, and $D_2$ denote total variation, Wasserstein-1 distance,
    Hellinger distance, Jensen--Shannon divergence, and Rényi divergence
    of order 2, respectively.}
  \label{si-fig:propagation-metric-comparison}
\end{figure}

\figref{si-fig:propagation-metric-comparison} compares RTA at the selected
interval widths with PACF across all four systems at
\qtylist{50;100;200}{\pico\second}.
At \qty{100}{\pico\second}, RTA has a lower discrepancy
on all six metrics in each SC system. POPC/water differs: PACF gives
lower total variation, Wasserstein-1 distance, Hellinger distance, and
Jensen--Shannon divergence, whereas RTA gives lower Rényi divergence of
order 2. PACF also gives a lower KL divergence, although the KL values
differ by only about \qty{5}{\percent}. At \qty{50}{\pico\second}, the SC/acetone
ranking also depends on the metric.

\FloatBarrier

\section{Permeability sensitivity to diffusivity uncertainty}

\label{si-sec:permeability}

The integration bounds are $x_a=-x_b$, with
$x_b=\qty{19}{\angstrom}$ for POPC/water and
$x_b=\qty{25.5}{\angstrom}$ for the SC systems. The POPC position was
estimated from the published
water-density profile; the SC position follows the earlier SC study.\cite{Thomas2026RTA,Thomas2025Skin}

To construct the pointwise-error sensitivity envelope, we apply the
fitting procedure in \secref{si-sec:profile-fitting} to the central estimates
$\widehat D_i$ and their pointwise limits
$\widehat D_i\pm1.96\,\operatorname{SE}(\widehat D_i)$.
We integrate all three profiles with the PMFs in
\secref{si-sec:reference-pmfs} over the bounds given above, using a grid
spacing of \qty{0.01}{\angstrom}.

Across all 120 system/width combinations, the fitted lower and upper
curves bracket the central diffusivity curve on the integration grid.
\tabref{si-tab:permeability-error}
reports the envelopes at the selected widths. They describe sensitivity
to the pointwise diffusivity errors and exclude PMF uncertainty.

\begin{table}[H]
  \centering
  \caption{Pointwise-error sensitivity envelope for RTA permeabilities
    at the selected interval widths, in \unit{\centi\metre\per\second},
    at the stated system-specific integration bounds. These envelopes are
    not \qty{95}{\percent} confidence intervals because spatial correlations
    of the diffusivity errors are unavailable.}
  \label{si-tab:permeability-error}
  \begin{tabular}{lrrrr}
    \toprule
    System     & $L$ / \unit{\angstrom} & Lower         & Central       & Upper         \\
    \midrule
    POPC/water & \num{12.5}             & \num{1.15e-3} & \num{1.61e-3} & \num{2.06e-3} \\
    SC/water   & \num{12}               & \num{1.44e-5} & \num{2.21e-5} & \num{2.97e-5} \\
    SC/acetone & \num{9.5}              & \num{4.05e-4} & \num{5.53e-4} & \num{6.97e-4} \\
    SC/6-MHO   & \num{11}               & \num{1.80e-2} & \num{2.39e-2} & \num{2.96e-2} \\
    \bottomrule
  \end{tabular}
\end{table}

\FloatBarrier

\section{POPC PMF sensitivity}

\label{si-sec:popc-pmf-sensitivity}

The PACF and VACF diffusivity estimates and unbiased MD propagators come
from the previously published 72-POPC system at
\qty{310}{\kelvin}.\cite{Thomas2026RTA} The new 100-POPC system used for
RTA differs in both size and temperature (\qty{308}{\kelvin}). The PMF
substitution below produces a modest change in permeability and preserves
the overall conclusions of the propagator comparison.

Only the PMF changes in this comparison; all other inputs are fixed.
Both PMFs are reflection-symmetrized, referenced to their mean over
$\qty{30}{\angstrom}\leq|x|\leq\qty{38}{\angstrom}$, and linearly
interpolated without smoothing.

\begin{figure}[!htbp]
  \centering
  \includegraphics[width=4.3in]{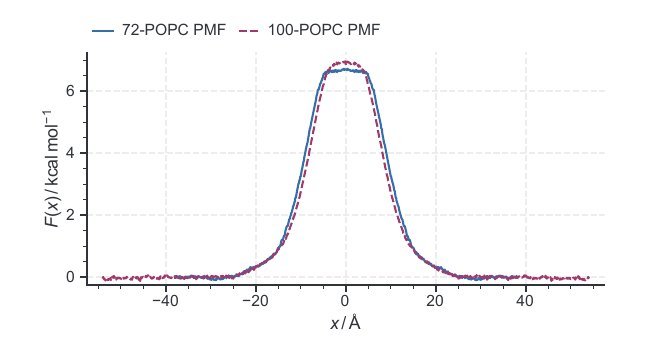}
  \caption{Water PMFs from the 72- and 100-POPC systems used in the
    sensitivity calculation.}
  \label{si-fig:popc-pmf-size-comparison}
\end{figure}

\begin{figure}[htbp]
  \centering
  \includegraphics[width=3.25in]{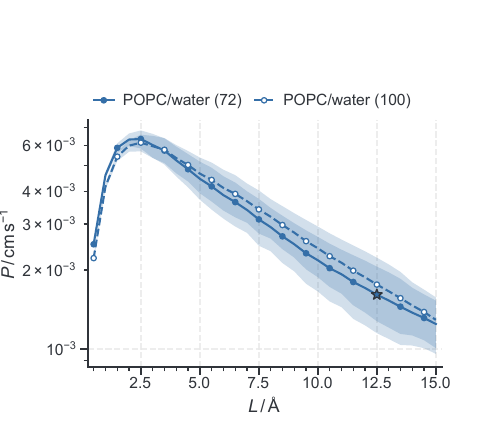}
  \caption{POPC/water permeability using the 72-POPC PMF (solid line,
    filled circles) and 100-POPC PMF (dashed line, open circles). The
    curves use the same RTA diffusivity profiles. Shading shows
    pointwise-error sensitivity envelopes; the star marks the selected
    width on the original curve.}
  \label{si-fig:popc-permeability-pmf-sensitivity}
\end{figure}

\figref{si-fig:popc-permeability-pmf-sensitivity} compares the permeability
curves across the interval-width scan. At the selected interval width
$L=\qty{12.5}{\angstrom}$,
permeability increases by \qty{8.2}{\percent}, from \qty{1.63e-3}{\centi\metre\per\second} to
\qty{1.76e-3}{\centi\metre\per\second}. This change is smaller than the
sensitivity to pointwise diffusivity errors at the selected width.

\begin{figure}[htbp]
  \centering
  \includegraphics[width=5.9in]{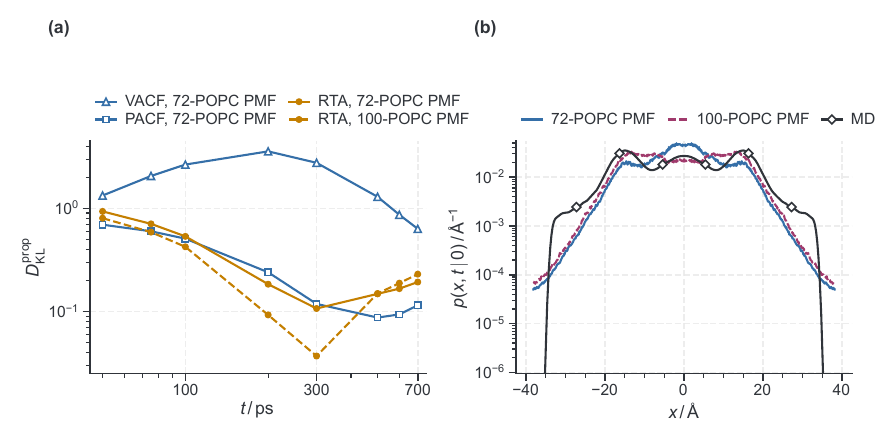}
  \caption{PMF sensitivity of RTA propagation at the selected interval width.
    (a) KL from the
    unbiased MD trajectories. RTA at $L=\qty{12.5}{\angstrom}$ uses the
    72-POPC (solid) and 100-POPC (dashed) PMFs; PACF and VACF use only
    the 72-POPC PMF.
    (b) RTA propagators at the selected interval width and
    $t=\qty{200}{\pico\second}$ compared with MD.}
  \label{si-fig:popc-pmf-propagator-sensitivity}
\end{figure}

\Needspace{5\baselineskip}
The PMF substitution reverses the RTA--PACF rankings at \qtylist{75;100}{\pico\second}
(\figref{si-fig:popc-pmf-propagator-sensitivity}). RTA at the selected interval
width remains closer to MD than VACF at every tested lag with either PMF.

\FloatBarrier

\section{Software}

\label{si-sec:software}

\begin{table}[H]
  \centering
  \caption{Software used for the new POPC simulations and figure preparation.
    Analysis and profile fitting use PDDA;\cite{PDDASoftware} numerical
    propagation uses SciPy.\cite{Virtanen2020SciPy}}
  \label{si-tab:software}
  \setlength{\tabcolsep}{5pt}
  \begin{tabular}{lll}
    \toprule
    Software                               & Version                 & Purpose                       \\
    \midrule
    CHARMM-GUI\cite{Jo2009MembraneBuilder} & ---                     & New POPC bilayer construction \\
    NAMD\cite{PhillipsHardy2020}           & 3.0.1                   & Molecular dynamics            \\
    Colvars\cite{FiorinKlein2013Colvars}   & Shipped with NAMD 3.0.1 & ABF sampling                  \\
    VMD\cite{HumphreyDalke1996VMD}         & 2.0.0a7                 & Molecular visualization       \\
    Tachyon\cite{Stone1998Tachyon}         & 0.99.5                  & Snapshot rendering            \\
    Matplotlib\cite{Hunter2007Matplotlib}  & 3.11.1                  & Plotting                      \\
    \bottomrule
  \end{tabular}
\end{table}

\clearpage\bibliography{references}

@article{Thomas2026RTA,
  author  = {Thomas, Rinto and Prabhakar, Praveen Ranganath and von Domaros, Michael},
  title   = {A Residence-Time Approach for Determining Position-Dependent Diffusivities from Biased Molecular Simulations},
  journal = {J. Chem. Theory Comput.},
  year    = {2026},
  volume  = {22},
  number  = {18},
  pages   = {9235--9246},
  doi     = {10.1021/acs.jctc.6c00645}
}

@article{Darve2001ABF,
  author  = {Darve, Eric and Pohorille, Andrew},
  title   = {Calculating Free Energies Using Average Force},
  journal = {J. Chem. Phys.},
  year    = {2001},
  volume  = {115},
  number  = {20},
  pages   = {9169--9183},
  doi     = {10.1063/1.1410978}
}

@article{Thomas2026SC,
  author  = {Thomas, Rinto and Prabhakar, Praveen Ranganath and Tobias, Douglas J. and von Domaros, Michael},
  title   = {Modeling Diffusion and Permeation Across the Stratum Corneum Lipid Barrier},
  journal = {ACS ES\&T Air},
  year    = {2026},
  volume  = {3},
  number  = {2},
  pages   = {580--589},
  doi     = {10.1021/acsestair.5c00412}
}

@article{Thomas2025Skin,
  author  = {Thomas, Rinto and Prabhakar, Praveen Ranganath and Tobias, Douglas J. and von Domaros, Michael},
  title   = {Insights into Dermal Permeation of Skin Oil Oxidation Products from Enhanced Sampling Molecular Dynamics Simulation},
  journal = {J. Phys. Chem. B},
  year    = {2025},
  volume  = {129},
  number  = {6},
  pages   = {1784--1794},
  doi     = {10.1021/acs.jpcb.4c08090}
}

@article{Diamond1974Partition,
  author  = {Diamond, J. M. and Katz, Y.},
  title   = {Interpretation of Nonelectrolyte Partition Coefficients between Dimyristoyl Lecithin and Water},
  journal = {J. Membr. Biol.},
  year    = {1974},
  volume  = {17},
  number  = {1},
  pages   = {121--154},
  doi     = {10.1007/BF01870176}
}

@article{Venable2019Permeability,
  author  = {Venable, Richard M. and Kr{\"a}mer, Andreas and Pastor, Richard W.},
  title   = {Molecular Dynamics Simulations of Membrane Permeability},
  journal = {Chem. Rev.},
  year    = {2019},
  volume  = {119},
  number  = {9},
  pages   = {5954--5997},
  doi     = {10.1021/acs.chemrev.8b00486}
}

@article{MarrinkBerendsen1994,
  author  = {Marrink, Siewert-Jan and Berendsen, Herman J. C.},
  title   = {Simulation of Water Transport through a Lipid Membrane},
  journal = {J. Phys. Chem.},
  year    = {1994},
  volume  = {98},
  number  = {15},
  pages   = {4155--4168},
  doi     = {10.1021/j100066a040}
}

@article{ComerChipot2013,
  author  = {Comer, Jeffrey and Chipot, Christophe and Gonz{\'a}lez-Nilo, Fernando D.},
  title   = {Calculating Position-Dependent Diffusivity in Biased Molecular Dynamics Simulations},
  journal = {J. Chem. Theory Comput.},
  year    = {2013},
  volume  = {9},
  number  = {2},
  pages   = {876--882},
  doi     = {10.1021/ct300867e}
}

@article{GhyselsVenable2017,
  author  = {Ghysels, An and Venable, Richard M. and Pastor, Richard W. and Hummer, Gerhard},
  title   = {Position-Dependent Diffusion Tensors in Anisotropic Media from Simulation: Oxygen Transport in and through Membranes},
  journal = {J. Chem. Theory Comput.},
  year    = {2017},
  volume  = {13},
  number  = {6},
  pages   = {2962--2976},
  doi     = {10.1021/acs.jctc.7b00039}
}

@article{ChipotComer2016,
  author  = {Chipot, Christophe and Comer, Jeffrey},
  title   = {Subdiffusion in Membrane Permeation of Small Molecules},
  journal = {Sci. Rep.},
  year    = {2016},
  volume  = {6},
  pages   = {35913},
  doi     = {10.1038/srep35913}
}

@article{KramerGhysels2020,
  author  = {Kr{\"a}mer, Andreas and Ghysels, An and Wang, Eric and Venable, Richard M. and Klauda, Jeffery B. and Brooks, Bernard R. and Pastor, Richard W.},
  title   = {Membrane Permeability of Small Molecules from Unbiased Molecular Dynamics Simulations},
  journal = {J. Chem. Phys.},
  year    = {2020},
  volume  = {153},
  number  = {12},
  pages   = {124107},
  doi     = {10.1063/5.0013429}
}

@article{GhyselsRoet2021,
  author  = {Ghysels, An and Roet, Sander and Davoudi, Samaneh and Van Erp, Titus S.},
  title   = {Exact Non-{Markovian} Permeability from Rare Event Simulations},
  journal = {Phys. Rev. Res.},
  year    = {2021},
  volume  = {3},
  number  = {3},
  pages   = {033068},
  doi     = {10.1103/PhysRevResearch.3.033068}
}

@article{HarrisChipotRoux2024,
  author  = {Harris, Jonathan and Chipot, Christophe and Roux, Beno{\^\i}t},
  title   = {Statistical Mechanical Theories of Membrane Permeability},
  journal = {J. Phys. Chem. B},
  year    = {2024},
  volume  = {128},
  number  = {38},
  pages   = {9183--9196},
  doi     = {10.1021/acs.jpcb.4c05020}
}

@article{Nagai2020LocalMSD,
  author  = {Nagai, Tetsuro and Tsurumaki, Shuhei and Urano, Ryo and Fujimoto, Kazushi and Shinoda, Wataru and Okazaki, Susumu},
  title   = {Position-Dependent Diffusion Constant of Molecules in Heterogeneous Systems as Evaluated by the Local Mean Squared Displacement},
  journal = {J. Chem. Theory Comput.},
  year    = {2020},
  volume  = {16},
  number  = {12},
  pages   = {7239--7254},
  doi     = {10.1021/acs.jctc.0c00448}
}

@article{GaalswykAwoonorWilliams2016,
  author  = {Gaalswyk, Kari and Awoonor-Williams, Ernest and Rowley, Christopher N.},
  title   = {Generalized {Langevin} Methods for Calculating Transmembrane Diffusivity},
  journal = {J. Chem. Theory Comput.},
  year    = {2016},
  volume  = {12},
  number  = {11},
  pages   = {5609--5619},
  doi     = {10.1021/acs.jctc.6b00747}
}

@book{Redner2001FirstPassage,
  author    = {Redner, Sidney},
  title     = {A Guide to First-Passage Processes},
  publisher = {Cambridge University Press},
  year      = {2001},
  doi       = {10.1017/CBO9780511606014}
}

@article{WoolfRoux1994,
  author  = {Woolf, Thomas B. and Roux, Benoit},
  title   = {Conformational Flexibility of O-Phosphorylcholine and O-Phosphorylethanolamine: A Molecular Dynamics Study of Solvation Effects},
  journal = {J. Am. Chem. Soc.},
  year    = {1994},
  volume  = {116},
  number  = {13},
  pages   = {5916--5926},
  doi     = {10.1021/ja00092a048}
}

@article{Hummer2005,
  author  = {Hummer, Gerhard},
  title   = {Position-Dependent Diffusion Coefficients and Free Energies from {Bayesian} Analysis of Equilibrium and Replica Molecular Dynamics Simulations},
  journal = {New J. Phys.},
  year    = {2005},
  volume  = {7},
  pages   = {34},
  doi     = {10.1088/1367-2630/7/1/034}
}

@article{ComerGumbart2015,
  author  = {Comer, Jeffrey and Gumbart, James C. and H{\'e}nin, J{\'e}r{\^o}me and Leli{\`e}vre, Tony and Pohorille, Andrew and Chipot, Christophe},
  title   = {The Adaptive Biasing Force Method: Everything You Always Wanted to Know but Were Afraid to Ask},
  journal = {J. Phys. Chem. B},
  year    = {2015},
  volume  = {119},
  number  = {3},
  pages   = {1129--1151},
  doi     = {10.1021/jp506633n}
}

@article{Jo2009MembraneBuilder,
  author  = {Jo, Sunhwan and Lim, Joseph B. and Klauda, Jeffery B. and Im, Wonpil},
  title   = {{CHARMM-GUI} Membrane Builder for Mixed Bilayers and Its Application to Yeast Membranes},
  journal = {Biophys. J.},
  year    = {2009},
  volume  = {97},
  number  = {1},
  pages   = {50--58},
  doi     = {10.1016/j.bpj.2009.04.013}
}

@article{KlaudaVenable2010,
  author  = {Klauda, Jeffery B. and Venable, Richard M. and Freites, J. Alfredo and O'Connor, Joseph W. and Tobias, Douglas J. and Mondragon-Ramirez, Carlos and Vorobyov, Igor and MacKerell, Alexander D. and Pastor, Richard W.},
  title   = {Update of the {CHARMM} All-Atom Additive Force Field for Lipids: Validation on Six Lipid Types},
  journal = {J. Phys. Chem. B},
  year    = {2010},
  volume  = {114},
  number  = {23},
  pages   = {7830--7843},
  doi     = {10.1021/jp101759q}
}

@article{VanommeslaegheHatcher2010,
  author  = {Vanommeslaeghe, Kenno and Hatcher, Elizabeth and Acharya, Chayan and Kundu, Shobhana and Zhong, Shijun and Shim, Jaehoon and Darian, Eric and Guvench, Olgun and Lopes, Pedro and Vorobyov, Igor and MacKerell, Alexander D.},
  title   = {{CHARMM} General Force Field: A Force Field for Drug-Like Molecules Compatible with the {CHARMM} All-Atom Additive Biological Force Fields},
  journal = {J. Comput. Chem.},
  year    = {2010},
  volume  = {31},
  number  = {4},
  pages   = {671--690},
  doi     = {10.1002/jcc.21367}
}

@article{JorgensenChandrasekhar1983,
  author  = {Jorgensen, William L. and Chandrasekhar, Jayaraman and Madura, Jeffry D. and Impey, Roger W. and Klein, Michael L.},
  title   = {Comparison of Simple Potential Functions for Simulating Liquid Water},
  journal = {J. Chem. Phys.},
  year    = {1983},
  volume  = {79},
  number  = {2},
  pages   = {926--935},
  doi     = {10.1063/1.445869}
}

@article{FlyvbjergPetersen1989,
  author  = {Flyvbjerg, H. and Petersen, H. G.},
  title   = {Error Estimates on Averages of Correlated Data},
  journal = {J. Chem. Phys.},
  year    = {1989},
  volume  = {91},
  number  = {1},
  pages   = {461--466},
  doi     = {10.1063/1.457480}
}

@article{Jonsson2018,
  author  = {Jonsson, Marius},
  title   = {Standard Error Estimation by an Automated Blocking Method},
  journal = {Phys. Rev. E},
  year    = {2018},
  volume  = {98},
  pages   = {043304},
  doi     = {10.1103/PhysRevE.98.043304}
}

@article{LiuHarderBerne2004,
  author  = {Liu, Pu and Harder, Edward and Berne, B. J.},
  title   = {On the Calculation of Diffusion Coefficients in Confined Fluids and Interfaces with an Application to the Liquid–Vapor Interface of Water},
  journal = {J. Phys. Chem. B},
  year    = {2004},
  volume  = {108},
  pages   = {6595--6602},
  doi     = {10.1021/jp0375057},
}

@article{MittalTruskett2008,
  author  = {Mittal, Jeetain and Truskett, Thomas M. and Errington, Jeffrey R. and Hummer, Gerhard},
  title   = {Layering and Position-Dependent Diffusive Dynamics of Confined Fluids},
  journal = {Phys. Rev. Lett.},
  year    = {2008},
  volume  = {100},
  pages   = {145901},
  doi     = {10.1103/PhysRevLett.100.145901},
}

@article{LauLubensky2007,
  author  = {Lau, A. W. C. and Lubensky, T. C.},
  title   = {State-dependent diffusion: Thermodynamic consistency and its path integral formulation},
  journal = {Phys. Rev. E},
  year    = {2007},
  volume  = {76},
  pages   = {011123},
  doi     = {10.1103/PhysRevE.76.011123},
}

@article{Kikkawa2026LocalDiffusion,
  author  = {Kikkawa, Nobuaki and Jinnouchi, Ryosuke and Nagai, Tetsuro and Okazaki, Susumu and Kimura, Masayuki},
  title   = {Local Diffusion Analysis Using Square Displacement Averaged in Subspace},
  journal = {J. Chem. Phys.},
  year    = {2026},
  volume  = {164},
  pages   = {114105},
  doi     = {10.1063/5.0304592},
}

@article{HarrisRoux2026,
  author  = {Harris, Jonathan and Roux, Benoît},
  title   = {Deep learning of position-dependent diffusivity from umbrella sampling molecular dynamics simulations},
  journal = {J. Chem. Phys.},
  year    = {2026},
  volume  = {165},
  pages   = {094118},
  doi     = {10.1063/5.0347525},
}

@article{MercierFrancoCastier2016,
  author  = {Mercier Franco, Lu{\'\i}s Fernando and Castier, Marcelo and Economou, Ioannis G.},
  title   = {Diffusion in homogeneous and in inhomogeneous media: A new unified approach},
  journal = {J. Chem. Theory Comput.},
  year    = {2016},
  volume  = {12},
  number  = {11},
  pages   = {5247--5255},
  doi     = {10.1021/acs.jctc.6b00653}
}

@article{HollringBaer2023,
  author  = {H{"o}llring, Kevin and Baer, Andreas and Vu{\v{c}}emilovi{\'c}-Alagi{\'c}, Nata{\v{s}}a and Smith, David M. and Smith, Ana-Sun{\v{c}}ana},
  title   = {Anisotropic molecular diffusion in confinement {I}: Transport of small particles in potential and density gradients},
  journal = {J. Colloid Interface Sci.},
  year    = {2023},
  volume  = {650},
  pages   = {1930--1940},
  doi     = {10.1016/j.jcis.2023.07.088}
}

@article{SicardKoskin2021,
  author  = {Sicard, Fran{\c{c}}ois and Koskin, Vladimir and Annibale, Alessia and Rosta, Edina},
  title   = {Position-dependent diffusion from biased simulations and {Markov} state model analysis},
  journal = {J. Chem. Theory Comput.},
  year    = {2021},
  volume  = {17},
  number  = {4},
  pages   = {2022--2033},
  doi     = {10.1021/acs.jctc.0c01151}
}

@article{Daldrop2017ExternalPotential,
  author  = {Daldrop, Jan O. and Kowalik, Bartosz G. and Netz, Roland R.},
  title   = {External Potential Modifies Friction of Molecular Solutes in Water},
  journal = {Phys. Rev. X},
  year    = {2017},
  volume  = {7},
  number  = {4},
  pages   = {041065},
  doi     = {10.1103/PhysRevX.7.041065}
}

@article{VervustZhang2023,
  author  = {Vervust, Wouter and Zhang, Daniel T. and Van Erp, Titus S. and Ghysels, An},
  title   = {Path sampling with memory reduction and replica exchange to reach long permeation timescales},
  journal = {Biophys. J.},
  year    = {2023},
  volume  = {122},
  number  = {14},
  pages   = {2960--2972},
  doi     = {10.1016/j.bpj.2023.02.021}
}

@article{KullbackLeibler1951,
  author  = {Kullback, Solomon and Leibler, Richard A.},
  title   = {On Information and Sufficiency},
  journal = {Ann. Math. Statist.},
  year    = {1951},
  volume  = {22},
  number  = {1},
  pages   = {79--86},
  doi     = {10.1214/aoms/1177729694}
}

@article{KucerkaTristramNagle2005,
  author  = {Kučerka, Norbert and Tristram-Nagle, Stephanie and Nagle, John F.},
  title   = {Structure of Fully Hydrated Fluid Phase Lipid Bilayers with Monounsaturated Chains},
  journal = {J. Membr. Biol.},
  year    = {2005},
  volume  = {208},
  number  = {3},
  pages   = {193--202},
  doi     = {10.1007/s00232-005-7006-8}
}

@article{WangKlauda2018SC,
  author  = {Wang, Eric and Klauda, Jeffery B.},
  title   = {Models for the Stratum Corneum Lipid Matrix: Effects of Ceramide Concentration, Ceramide Hydroxylation, and Free Fatty Acid Protonation},
  journal = {J. Phys. Chem. B},
  year    = {2018},
  volume  = {122},
  number  = {50},
  pages   = {11996--12008},
  doi     = {10.1021/acs.jpcb.8b06188}
}

@article{WisthalerWeschler2010,
  author  = {Wisthaler, Armin and Weschler, Charles J.},
  title   = {Reactions of Ozone with Human Skin Lipids: Sources of Carbonyls, Dicarbonyls, and Hydroxycarbonyls in Indoor Air},
  journal = {Proc. Natl. Acad. Sci. U.S.A.},
  year    = {2010},
  volume  = {107},
  number  = {15},
  pages   = {6568--6575},
  doi     = {10.1073/pnas.0904498106}
}

@article{vonDomaros2020Multiscale,
  author  = {von Domaros, Michael and Lakey, Pascale S. J. and Shiraiwa, Manabu and Tobias, Douglas J.},
  title   = {Multiscale Modeling of Human Skin Oil-Induced Indoor Air Chemistry: Combining Kinetic Models and Molecular Dynamics},
  journal = {J. Phys. Chem. B},
  year    = {2020},
  volume  = {124},
  number  = {18},
  pages   = {3836--3843},
  doi     = {10.1021/acs.jpcb.0c02818}
}

@article{Lakey2019SkinClothing,
  author  = {Lakey, Pascale S. J. and Morrison, Glenn C. and Won, Youngbo and Parry, Krista M. and von Domaros, Michael and Tobias, Douglas J. and Rim, Donghyun and Shiraiwa, Manabu},
  title   = {The Impact of Clothing on Ozone and Squalene Ozonolysis Products in Indoor Environments},
  journal = {Commun. Chem.},
  year    = {2019},
  volume  = {2},
  pages   = {56},
  doi     = {10.1038/s42004-019-0159-7}
}

@article{Shiraiwa2019MOCCIE,
  author  = {Shiraiwa, Manabu and Carslaw, Nicola and Tobias, Douglas J. and Waring, Michael S. and Rim, Donghyun and Morrison, Glenn and Lakey, Pascale S. J. and Kruza, Magdalena and von Domaros, Michael and Cummings, Bryan E. and Won, Youngbo},
  title   = {Modelling Consortium for Chemistry of Indoor Environments ({MOCCIE}): Integrating Chemical Processes from Molecular to Room Scales},
  journal = {Environ. Sci.: Processes Impacts},
  year    = {2019},
  volume  = {21},
  number  = {8},
  pages   = {1240--1254},
  doi     = {10.1039/C9EM00123A}
}

@misc{PDDASoftware,
  author       = {von Domaros, Michael},
  title        = {{PDDA}: Position-Dependent Diffusivity Analysis},
  howpublished = {GitHub repository},
  year         = {2026},
  url          = {https://github.com/mvondomaros-lab/pdda}
}

@article{PhillipsHardy2020,
  author  = {Phillips, James C. and Hardy, David J. and Maia, Julio D. C. and Stone, John E. and Ribeiro, Jo{\~a}o V. and Bernardi, Rafael C. and Buch, Ronak and Fiorin, Giacomo and H{\'e}nin, J{\'e}r{\^o}me and Jiang, Wei and McGreevy, Ryan and Melo, Marcelo C. R. and Radak, Brian K. and Skeel, Robert D. and Singharoy, Abhishek and Wang, Yi and Roux, Beno{\^\i}t and Aksimentiev, Aleksei and Luthey-Schulten, Zaida and Kal{\'e}, Laxmikant V. and Schulten, Klaus and Chipot, Christophe and Tajkhorshid, Emad},
  title   = {Scalable molecular dynamics on {CPU} and {GPU} architectures with {NAMD}},
  journal = {J. Chem. Phys.},
  year    = {2020},
  volume  = {153},
  number  = {4},
  pages   = {044130},
  doi     = {10.1063/5.0014475}
}

@article{FiorinKlein2013Colvars,
  author  = {Fiorin, Giacomo and Klein, Michael L. and H{\'e}nin, J{\'e}r{\^o}me},
  title   = {Using Collective Variables to Drive Molecular Dynamics Simulations},
  journal = {Mol. Phys.},
  year    = {2013},
  volume  = {111},
  number  = {22--23},
  pages   = {3345--3362},
  doi     = {10.1080/00268976.2013.813594}
}

@article{HumphreyDalke1996VMD,
  author  = {Humphrey, William and Dalke, Andrew and Schulten, Klaus},
  title   = {{VMD}: Visual Molecular Dynamics},
  journal = {J. Mol. Graph.},
  year    = {1996},
  volume  = {14},
  number  = {1},
  pages   = {33--38},
  doi     = {10.1016/0263-7855(96)00018-5}
}

@mastersthesis{Stone1998Tachyon,
  author = {Stone, John E.},
  title  = {An Efficient Library for Parallel Ray Tracing and Animation},
  school = {University of Missouri--Rolla},
  year   = {1998}
}

@article{Hunter2007Matplotlib,
  author  = {Hunter, John D.},
  title   = {{Matplotlib}: A 2D Graphics Environment},
  journal = {Comput. Sci. Eng.},
  year    = {2007},
  volume  = {9},
  number  = {3},
  pages   = {90--95},
  doi     = {10.1109/MCSE.2007.55}
}

@article{FellerZhang1995Piston,
  author  = {Feller, Scott E. and Zhang, Yuhong and Pastor, Richard W. and Brooks, Bernard R.},
  title   = {Constant Pressure Molecular Dynamics Simulation: The {Langevin} Piston Method},
  journal = {J. Chem. Phys.},
  year    = {1995},
  volume  = {103},
  number  = {11},
  pages   = {4613--4621},
  doi     = {10.1063/1.470648}
}

@article{BussiDonadio2007,
  author  = {Bussi, Giovanni and Donadio, Davide and Parrinello, Michele},
  title   = {Canonical Sampling through Velocity Rescaling},
  journal = {J. Chem. Phys.},
  year    = {2007},
  volume  = {126},
  number  = {1},
  pages   = {014101},
  doi     = {10.1063/1.2408420}
}

@article{TuckermanBerne1992,
  author  = {Tuckerman, Mark and Berne, Bruce J. and Martyna, Glenn J.},
  title   = {Reversible Multiple Time Scale Molecular Dynamics},
  journal = {J. Chem. Phys.},
  year    = {1992},
  volume  = {97},
  number  = {3},
  pages   = {1990--2001},
  doi     = {10.1063/1.463137}
}

@article{MiyamotoKollman1992SETTLE,
  author  = {Miyamoto, Shuichi and Kollman, Peter A.},
  title   = {{SETTLE}: An Analytical Version of the {SHAKE} and {RATTLE} Algorithm for Rigid Water Models},
  journal = {J. Comput. Chem.},
  year    = {1992},
  volume  = {13},
  number  = {8},
  pages   = {952--962},
  doi     = {10.1002/jcc.540130805}
}

@article{RyckaertCiccotti1977SHAKE,
  author  = {Ryckaert, Jean-Paul and Ciccotti, Giovanni and Berendsen, Herman J. C.},
  title   = {Numerical Integration of the Cartesian Equations of Motion of a System with Constraints: Molecular Dynamics of n-Alkanes},
  journal = {J. Comput. Phys.},
  year    = {1977},
  volume  = {23},
  number  = {3},
  pages   = {327--341},
  doi     = {10.1016/0021-9991(77)90098-5}
}

@article{Essmann1995PME,
  author  = {Essmann, Ulrich and Perera, Lalith and Berkowitz, Max L. and Darden, Tom and Lee, Hsing and Pedersen, Lee G.},
  title   = {A Smooth Particle Mesh {Ewald} Method},
  journal = {J. Chem. Phys.},
  year    = {1995},
  volume  = {103},
  number  = {19},
  pages   = {8577--8593},
  doi     = {10.1063/1.470117}
}

@article{RousseeuwCroux1993MAD,
  author  = {Rousseeuw, Peter J. and Croux, Christophe},
  title   = {Alternatives to the Median Absolute Deviation},
  journal = {J. Am. Stat. Assoc.},
  year    = {1993},
  volume  = {88},
  number  = {424},
  pages   = {1273--1283},
  doi     = {10.1080/01621459.1993.10476408}
}

@book{Wahba1990Splines,
  author    = {Wahba, Grace},
  title     = {Spline Models for Observational Data},
  publisher = {Society for Industrial and Applied Mathematics},
  address   = {Philadelphia, PA},
  year      = {1990},
  series    = {CBMS-NSF Regional Conference Series in Applied Mathematics},
  volume    = {59},
  doi       = {10.1137/1.9781611970128}
}

@article{GibbsSu2002Metrics,
  author  = {Gibbs, Alison L. and Su, Francis Edward},
  title   = {On Choosing and Bounding Probability Metrics},
  journal = {Int. Stat. Rev.},
  year    = {2002},
  volume  = {70},
  number  = {3},
  pages   = {419--435},
  doi     = {10.1111/j.1751-5823.2002.tb00178.x}
}

@article{Lin1991JS,
  author  = {Lin, Jianhua},
  title   = {Divergence Measures Based on the {Shannon} Entropy},
  journal = {IEEE Trans. Inf. Theory},
  year    = {1991},
  volume  = {37},
  number  = {1},
  pages   = {145--151},
  doi     = {10.1109/18.61115}
}

@article{VanErvenHarremoes2014,
  author  = {van Erven, Tim and Harremo{\"e}s, Peter},
  title   = {{R\'enyi} Divergence and {Kullback--Leibler} Divergence},
  journal = {IEEE Trans. Inf. Theory},
  year    = {2014},
  volume  = {60},
  number  = {7},
  pages   = {3797--3820},
  doi     = {10.1109/TIT.2014.2320500}
}

@article{WangPeskinElston2003,
  author = {Wang, Hongyun and Peskin, Charles S. and Elston, Timothy C.},
  title = {A Robust Numerical Algorithm for Studying Biomolecular Transport Processes},
  journal = {J. Theor. Biol.},
  year = {2003},
  volume = {221},
  number = {4},
  pages = {491--511},
  doi = {10.1006/jtbi.2003.3200}
}

@article{AlMohyHigham2011,
  author = {Al-Mohy, Awad H. and Higham, Nicholas J.},
  title = {Computing the Action of the Matrix Exponential, with an Application to Exponential Integrators},
  journal = {SIAM J. Sci. Comput.},
  year = {2011},
  volume = {33},
  number = {2},
  pages = {488--511},
  doi = {10.1137/100788860}
}

@article{DeVosVenable2018,
  author  = {De Vos, Oriana and Venable, Richard M. and Van Hecke, Tanja and Hummer, Gerhard and Pastor, Richard W. and Ghysels, An},
  title   = {Membrane Permeability: Characteristic Times and Lengths for Oxygen, and a Simulation-Based Test of the Inhomogeneous Solubility-Diffusion Model},
  journal = {J. Chem. Theory Comput.},
  year    = {2018},
  volume  = {14},
  number  = {7},
  pages   = {3811--3824},
  doi     = {10.1021/acs.jctc.8b00115}
}

@article{Paninski2003InformationEstimation,
  author  = {Paninski, Liam},
  title   = {Estimation of Entropy and Mutual Information},
  journal = {Neural Comput.},
  year    = {2003},
  volume  = {15},
  number  = {6},
  pages   = {1191--1253},
  doi     = {10.1162/089976603321780272}
}

@article{Virtanen2020SciPy,
  author  = {Virtanen, Pauli and Gommers, Ralf and Oliphant, Travis E. and Haberland, Matt and Reddy, Tyler and Cournapeau, David and Burovski, Evgeni and Peterson, Pearu and Weckesser, Warren and Bright, Jonathan and van der Walt, St{\'e}fan J. and Brett, Matthew and Wilson, Joshua and Millman, K. Jarrod and Mayorov, Nikolay and Nelson, Andrew R. J. and Jones, Eric and Kern, Robert and Larson, Eric and Carey, C. J. and Polat, {\.{I}}lhan and Feng, Yu and Moore, Eric W. and VanderPlas, Jake and Laxalde, Denis and Perktold, Josef and Cimrman, Robert and Henriksen, Ian and Quintero, E. A. and Harris, Charles R. and Archibald, Anne M. and Ribeiro, Ant{\^o}nio H. and Pedregosa, Fabian and van Mulbregt, Paul and {SciPy 1.0 Contributors}},
  title   = {{SciPy} 1.0: Fundamental Algorithms for Scientific Computing in {Python}},
  journal = {Nat. Methods},
  year    = {2020},
  volume  = {17},
  number  = {3},
  pages   = {261--272},
  doi     = {10.1038/s41592-019-0686-2}
}
\end{document}